\documentclass[preprint]{elsarticle}

\usepackage{lineno,hyperref}
\usepackage[procnames]{listings}

\usepackage{color}

\modulolinenumbers[5]

\usepackage{ulem}

\journal{Journal of \LaTeX\ Templates}

\begin{document}
\definecolor{keywords}{RGB}{255,0,90}
\definecolor{comments}{RGB}{0,0,113}
\definecolor{red}{RGB}{160,0,0}
\definecolor{green}{RGB}{0,150,0}
 
\renewcommand{\linenumbers}{}

\begin{frontmatter}

\title{Multi Order Coverage data structure to plan multi-messenger observations}
\author{Giuseppe Greco\fnref{myfootnote}}
\author{Michele Punturo}
\address{INFN, Sezione di Perugia, I-06123 Perugia, Italy}
\author{Mark Allen, Ada Nebot, Pierre Fernique,  Matthieu Baumann, Fran\c{c}ois-Xavier Pineau, Thomas Boch, S\'{e}bastien Derriere}
\address{Universit\'{e} de Strasbourg, CNRS, Observatoire astronomique de Strasbourg, UMR 7550, F-67000, Strasbourg, France}
\author{Marica Branchesi}
\address{Gran Sasso Science Institute (GSSI), Aquila, Italy}
\address{INFN, Laboratori Nazionali del Gran Sasso, I-67100 Assergi, Italy}
\author{Mateusz Bawaj, Helios Vocca}
\address{Universit\`{a} degli Studi di Perugia, Perugia, Italy}
\address{INFN - Sezione di Perugia, Italy}
\fntext[myfootnote]{giuseppe.greco@pg.infn.it}

\begin{abstract}
We describe the use of Multi Order Coverage (MOC) maps as a practical way to manage complex regions of the sky for the planning of multi-messenger observations. MOC maps are a data structure that provides a multi-resolution representation of irregularly shaped and fragmentary regions over the sky based on the HEALPix (Hierarchical Equal Area isoLatitude Pixelization) tessellation. We present a new application of MOC, in combination with the \texttt{astroplan} observation planning package, to enable the efficient computation of sky regions and the visibility of these regions from a specific location on the Earth at a particular time. 

Using the example of the low-latency gravitational-wave alerts, and a simulated observational campaign with three observatories, we show that the use of MOC maps allows a high level of interoperability to support observing schedule plans. Gravitational-wave detections have an associated credible region localization on the sky. We demonstrate that these localizations can be encoded as MOC maps, and how they can be used in
visualisation tools, and processed (filtered, combined) and also their utility for access to Virtual Observatory services which can be queried 'by MOC' for data within the region of interest. The ease of generating the MOC maps and the fast access to data means that the whole system can be very efficient, so that any updates on the gravitational-wave sky localization can be quickly taken into account and the corresponding adjustments to observing schedule plans can be rapidly implemented. We provide example python code as a practical example of these methods. In addition, a video demonstration of the entire  workflow is available.
\end{abstract}

\begin{keyword}
\texttt{multi-messenger, gravitational waves, multi order coverage map, sky localization, visibility}
\end{keyword}
\end{frontmatter}
\linenumbers

\section{Multi-messenger data science tools and worldwide collaborations}
The first discovery of gravitational waves from the coalescence of two neutron stars in a binary system, GW170817, was made by the LIGO and Virgo collaboration (LVC) \cite{gw170817} during the second observational run of LVC \cite{mma_lvc}. The subsequent identification of its electromagnetic counterpart, GRB170817 / AT2017gfo, has ushered the science community in the era of multi-messenger astronomy \cite{gw170817_mma}. 

During the LVC third observing run new data science tools have been developed, and worldwide collaborations have been organised to support coordinated follow-up observations and their analysis, see \cite{nature_astronomy} and references therein.
The future growth of multi-messenger astronomy that is expected in the next decades \cite{observing_paper}, \cite{Maggiore_2020} will require an appreciable coordination effort and a significant synergy between different ground-based observatories, space-borne satellites and second/third-generation interferometric gravitational wave detectors (Virgo \cite{virgo}, LIGO \cite{ligo}, KAGRA \cite{kagra}, Cosmic Explorer \cite{cosmic_explorer}, Einstein Telescope (ET) \cite{et}).
 
The establishment of the interoperable infrastructures that will be needed to support observatory operations, astronomical research, and data storage,
are achievable goals if the active multi-messenger community cooperates in the development of  technical standards and common protocols. The International Virtual Observatory Alliance (IVOA) \cite{ivoa} is an organisation that develops the technical standards to lay the foundation of the astronomical Virtual Observatory (VO) \cite{vo}, \cite{vo2}, \cite{vo3}. 
A number of VO-based protocols are being developed and updated to respond to the needs of multi-messenger astronomy. For example, two of the protocols
in development for the planning of multi-messenger observations \cite{ness} are:  $(i)$ the Observation Locator Table Access Protocol (ObsLocTAP)\footnote{\url{https://github.com/ivoa-std/ObsLocTAP}}
to define a data model for observation schedules,  and $(ii)$
the Object Visibility Simple Access Protocol  (ObjVisSAP)\footnote{\url{https://github.com/ivoa-std/ObjVisSAP}}  for retrieving the visibility of astronomical objects (based on their sky coordinates) through  a uniform query interface. These protocols are expected to be very useful for the planning of observations in the cases where object coordinates are known, or where the planning can be done point by point.

In this article, we consider the cases where astrophysical sources may be localised in relatively large sky regions, and in particular the case of multi-messenger observation campaigns for gravitational wave follow-ups.

We describe a practical approach that uses Multi-Order Coverage (MOC) maps \cite{moc1.1} to describe regions on the sky that are observable from given locations on the Earth, taking into account specific constraints on the airmass values and the time allocated for the observation. The identification of existing image data is an integral part of this process, which also benefits significantly from the MOC based approach. Operationally, this approach uses Python modules for the creation 
and manipulation of MOCs (\texttt{mocpy}; \cite{mocpy19}, \cite{mocpy20}) and for use of the HEALPix tessellation \cite{Gorski_2005} (\texttt{cds-healpix-python}; \cite{cds_healpix19}), combined with the  \texttt{astroplan} 
package \cite{astroplan2018} to select the HEALPix indices at a given order and
to set up a new MOC map that represents the visibility with all of these constraints taken into account. Note that the complete code is reported in a public GitHub repository$\footnote{\url{https://github.com/ggreco77/MOC-to-plan-MMA}}$  
which indicates all of the Python modules necessary for this analysis.
A video demonstration$\footnote{\url{https://virgo.pg.infn.it/sites/virgo.pg.infn.it/files/mm/tuto_A_C_mid.mp4}}$ is also provided focusing on the Aladin functionalities used through the paper.

\section{Multi Order Coverage Map (MOC)}
The MOC encoding method was originally developed at the Centre de Donn\'{e}es astronomiques de Strasbourg (CDS) and has been adopted as a recommendation 
by the IVOA as the MOC 1.0 standard \citep{fernique14} and updated with the MOC 1.1 standard \citep{moc1.1}. Initially designed for manipulating sky coverages from astronomical surveys, MOC is currently being extended  to  support both  temporal and spatial coverage in the MOC 2.0 standard \cite{fernique21} that is currently at the "Proposed Recommendation" status in the IVOA standardisation process. MOC 2.0 considers space MOCs (SMOC), time MOCs (TMOC) and the combined  space-Time MOC (STMOC). 

The ability to transform any spatial MOC generated from telescope footprints, image surveys, catalogues, gravitational-wave sky localizations, MOC visibility etc. into  spatial and temporal MOC (STMOC) \cite{fernique21} could have an impact both in  immediate collaboration and sharing between different networks and in archival searches to investigate the current theoretical models (see \textbf{Fig. 4} in  \cite{fernique21}).
This paper concerns mainly space MOCs (which we refer to simply as MOCs) with a small application of the STMOC. 

MOCs are a very useful mechanism to improve the efficiency of querying VO services. A number of services support queries based on MOCs in different ways. For example the CDS VizieR service allows catalogues of astronomical sources to be queried by MOC to provide a list of sources within the MOC. Another example is the use of MOCs in the 20 HiPS node sites, where the MOC file (as described in the HiPS standard - \cite{hips_ivoa}) can be used to avoid loading any HiPS tiles located outside the coverage of the HiPS data. Furthermore the CDS MOC Server\footnote{\url{https://alasky.unistra.fr/MocServer/query}} includes the MOCs of  $\sim27000$ image and catalogue data sets, which can help by providing a way to limit queries only to data sets that have coverages in the sky region of interest. Such considerations help optimise the response times of data searches which can be critical in planning of observations.

\subsection{A brief description of MOC} 
The MOC data structure is based on the HEALPix 
(\textbf{H}ierarchical \textbf{E}qual \textbf{A}rea iso\textbf{L}atitude \textbf{Pix}elization)  sky tessellation algorithm \cite{Gorski_2005}. 
In its base partitioning, a HEALPix sphere is hierarchically tessellated into curvilinear quadrilaterals.
The lowest resolution partition consists of twelve base pixels. 
The resolution of the tessellation, indicated by the parameter $nside$,  increases by division of each pixel into four new ones. 
The number of pixels composing the sphere at a specific resolution is $12\times nside^2$.

The fundamental element of a MOC data structure is the HEALPix \textit{cell} which uniquely defines  a region on the sky.
Each MOC cell is defined by two HEALPix properties: the hierarchy level (HEALPix $order$) and the pixel index (HEALPix $npix$). 
The finest level of refinement within the MOC hierarchy is determined by the HEALPix order parameter or the equivalent $nside$;
$order$ and $nside$ are related by

\begin{equation}
nside=2^{order}. 
\label{eq:nside_order}
\end{equation}
For example, $order = 9$ corresponds to $nside = 512$, 
and a resolution of 6.9' per pixel.

\begin{figure}
	\includegraphics[width=\linewidth]{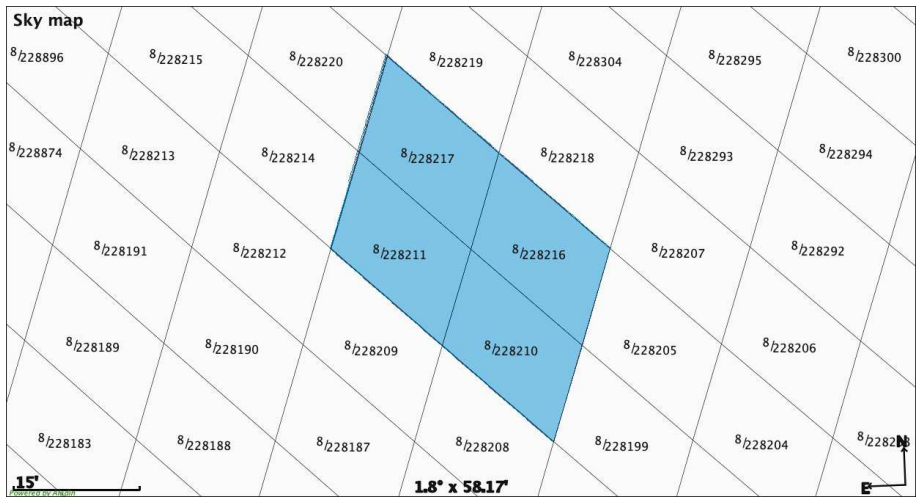}
	\caption{HEALPix grid. The shaded area highlights a MOC map consisting 
	         of 4 cells at $order$ = 8. Lower-order cells along the edges 
	         are not considered here.}
	\label{fig:healpix_grid}
\end{figure}

\subsection{Serialisations} 
\label{subsec:serialisations}
Two main serialisations of a MOC are supported: JSON and  FITS. Serialisations are defined in the MOC standard document \cite{fernique14} (see sections {\bf 2.3.2} (FITS) 
and {\bf 3.1.1} (JSON))

A JSON MOC is written following the syntax:

\begin{equation}
          \{  "order":[npix,npix,...], "order":[npix, npix...], ... \}.
\end{equation}

In the example shown in Figure \ref{fig:healpix_grid}, the JSON MOC shown as the shaded region is defined by 4 cells at order = 8:

\begin{equation}
          \{"8":[228210,228211,228216,228217]\}.
\end{equation}

To encode a MOC in a FITS file, each cell is converted into a single integer using the standard NUNIQ packing scheme (see \cite{fernique14}):

\begin{equation}
          uniq = 4 \times (4^{order}) + npix
\end{equation}

In Figure \ref{fig:healpix_grid}, the FITS MOC contains the integers \texttt{490354, 490355, 490360, 490361}, 
and the resulting list is stored in a single column binary table extension. 
The UNIQUE constraint ensures that all values in the column are different.
For $order$ $<=13$ these UNIQUE values can be stored in a 32-bit signed integer, and for $order$ $<=$29 in a 64-bit signed integer.
 
\begin{figure}
\center
	\includegraphics[width=6cm]{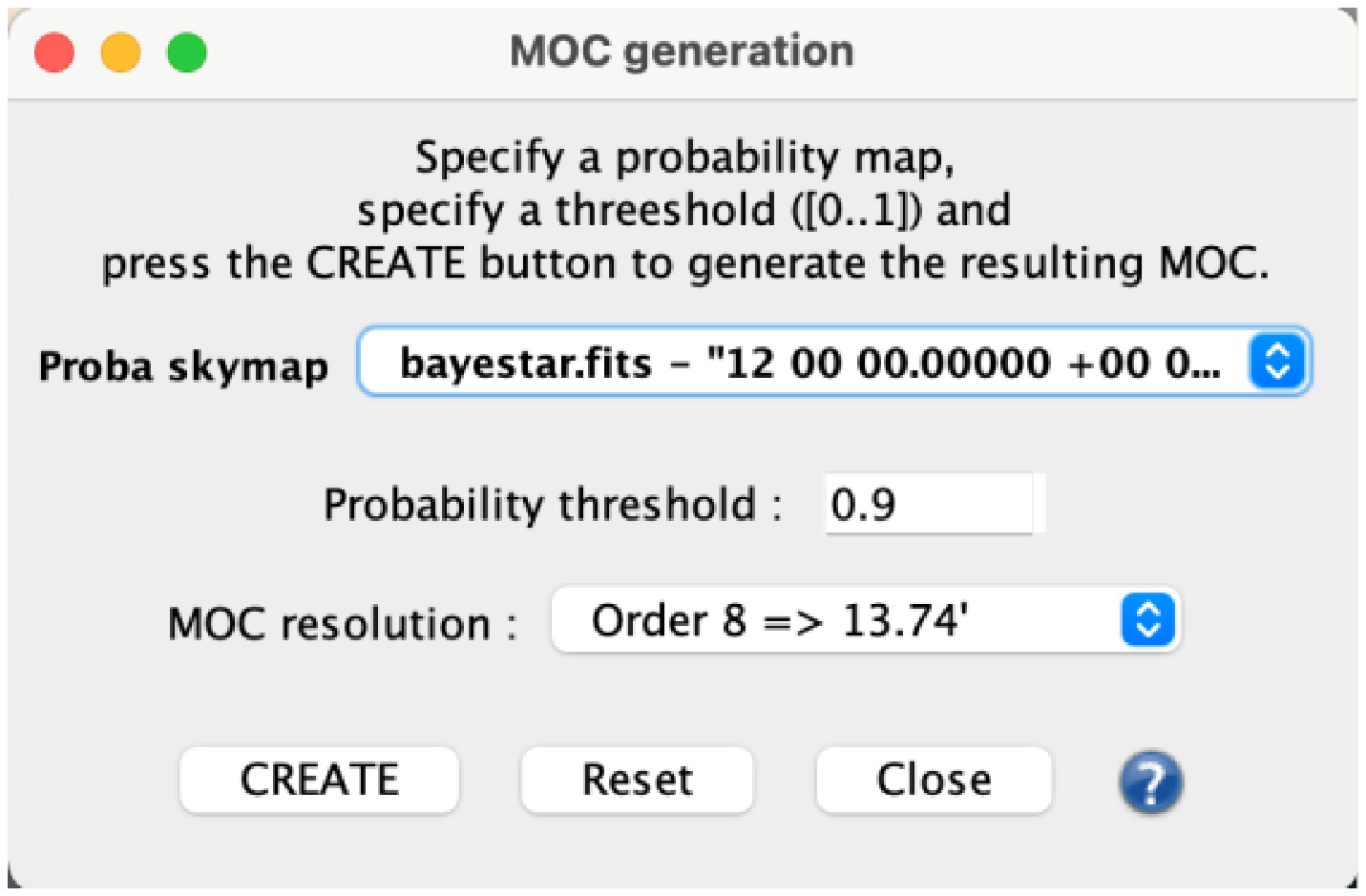}
	\includegraphics[width=6cm]{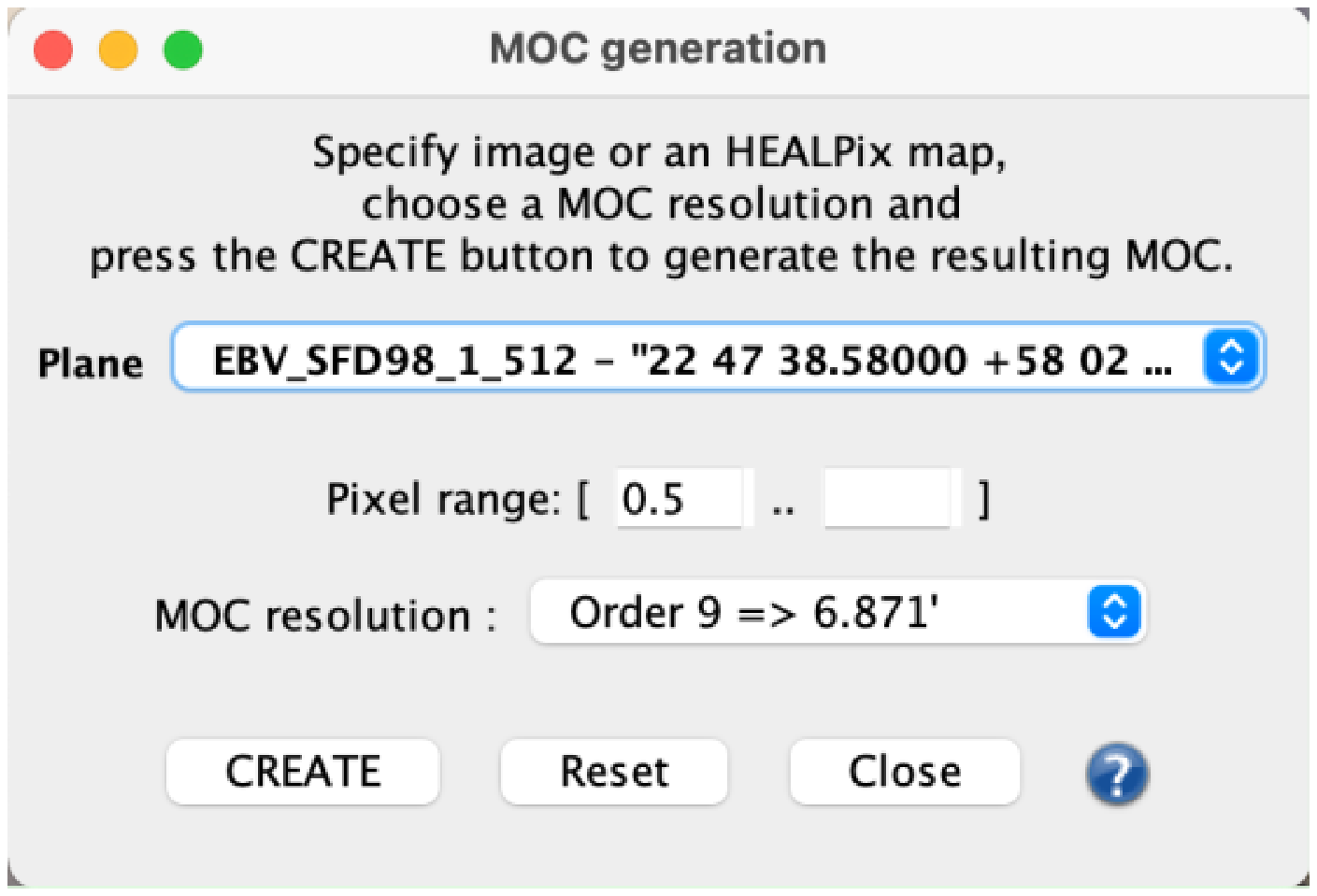}
	\caption{ MOC generation windows. The two main Aladin windows which are used to encode a sky region in a MOC data structure. \textit{Left}: from a probability threshold. \textit{Right}: from a collection of pixel values. The sky maps are selected from the dropdown menu \textbf{Proba skymap} and \textbf{Plane}, respectively. In both windows, the MOC order can be chosen from the dropdown menu \textbf{MOC resolution}. }
	\label{fig:window_moc}
\end{figure}

\begin{figure}
\centering
	\includegraphics[width=0.96\textwidth]{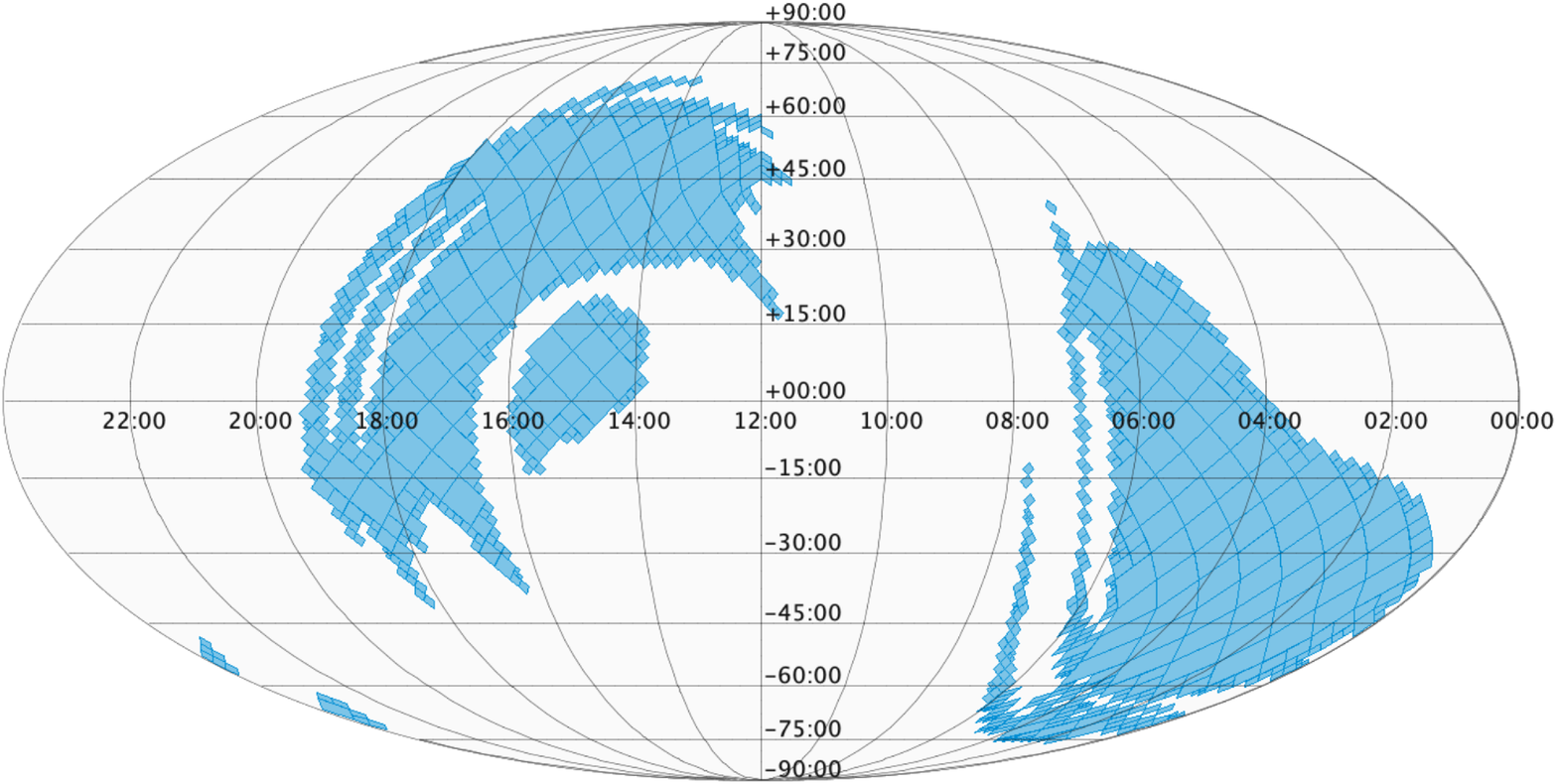}
	\includegraphics[width=0.96\textwidth]{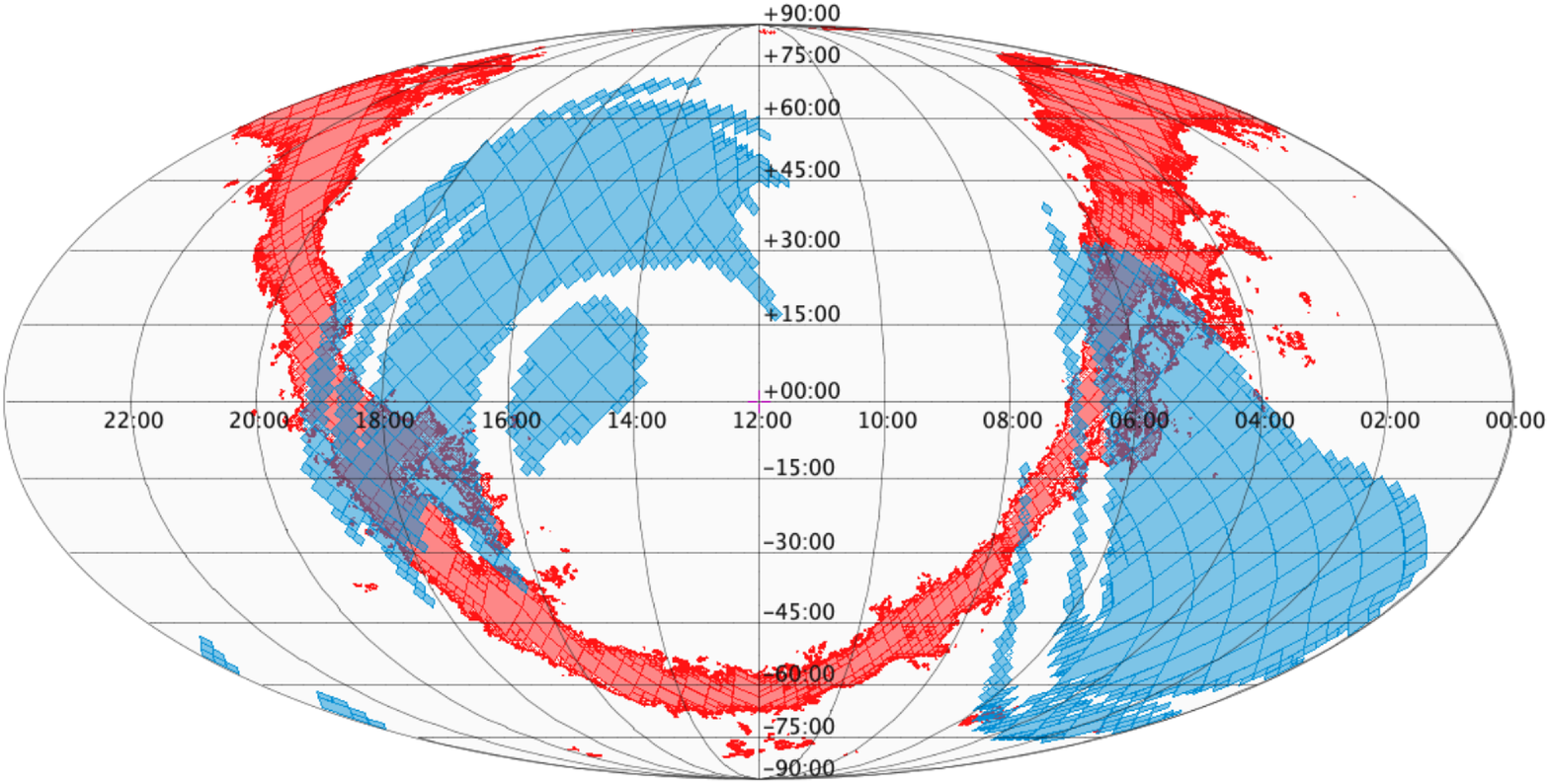}
	\includegraphics[width=0.96\textwidth]{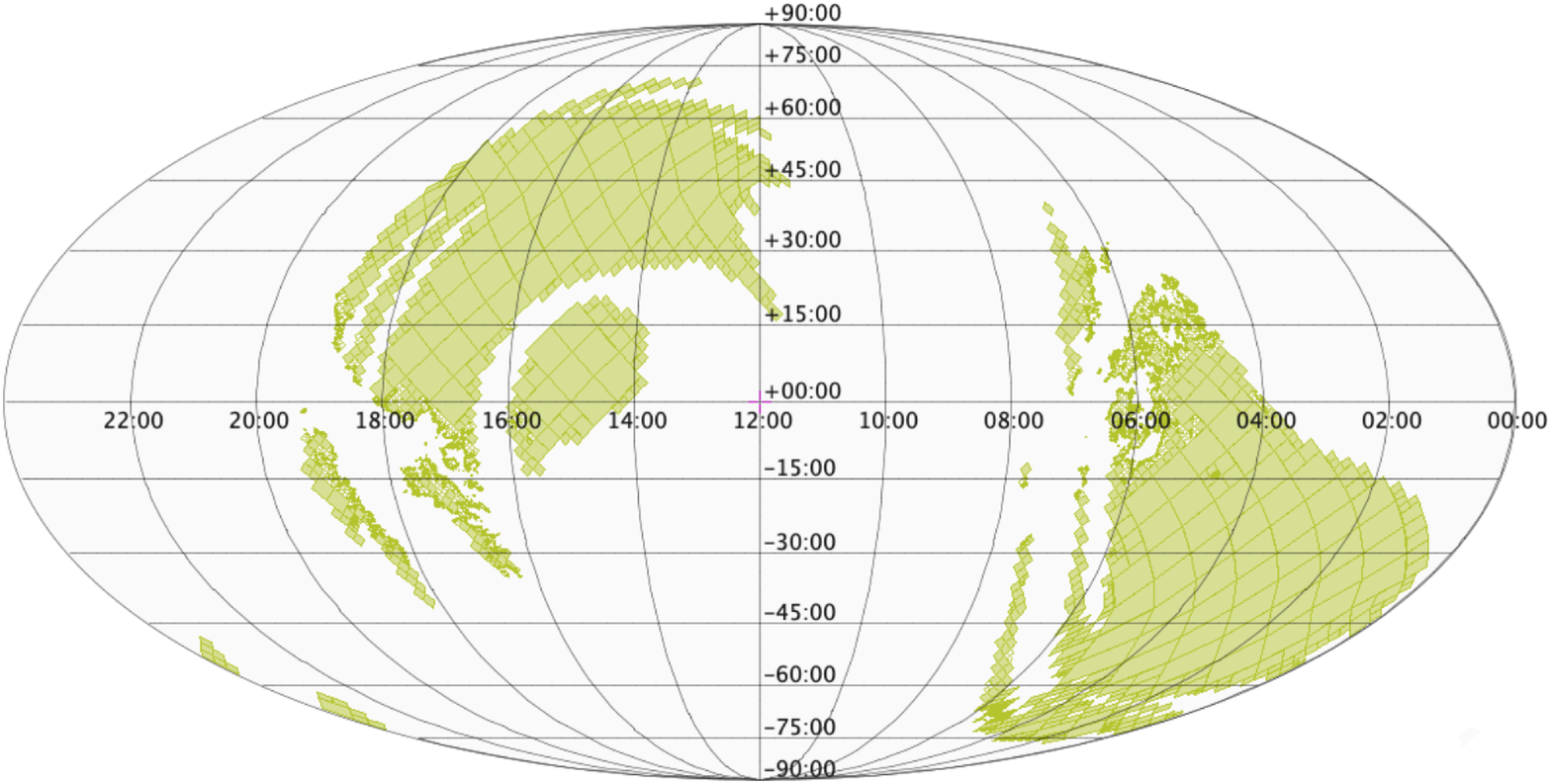}
	\caption{\textit{Top:}  
	         Gravitational-wave sky localization of GW190425 - 90\% credible region -- Mollweide projection.
	         \textit{Middle:}   
	         A selected high Galactic dust extinction region is overlaid. The range of pixel values was set from 0.5 up to the highest value.  
	         \textit{Bottom:}  
	         The resulting sky map defined by the gravitational-wave sky localization with the dust absorption fields subtracted.}
	\label{fig:skymap_operations_1}
\end{figure}

\section{Encoding of 2D credible regions in a MOC map}
\label{sec:encoding}
The 2D credible regions of  gravitational-wave sky localizations can be straightforwardly encoded into a MOC data structure \cite{greco2019}, \cite{greco_adass}.
The contours of a gravitational-wave sky localization are constructed following the prescription in \citep{singer14}: we ranked the pixels from most probable to least, and
finally counted how many pixels summed to a given total probability. 
The algorithm then constructs the MOC coverage  extracting the HEALPix pixels inside 
that contour plot.
Every single level of the contained probability can be used as a regular MOC even when the sky localization is irregularly shaped with disjoint regions. 

Alternatively, the credible regions can be described by a polygon enclosing a certain percentage of probability. This approach returns an output in which each point is represented by its right ascension and declination and where special terminators are added to delimit disjointed regions if needed. The MOC data structure is more compact, returning a flat list of integers (in a FITS serialisation) and managing the HEALPix resolution ($nside$) by the MOC order parameter. 

Here, the excellent performances of the MOC maps are shown in the context of the multi-messenger science in which data visualisation, data access and data comparison are crucial tasks to search for electromagnetic counterparts from gravitational-wave sources.

In the next section, we demonstrate two tools that can be used to fulfil the task to build MOC data structures from a gravitational-wave sky localization: Aladin Desktop \cite{aladin2000} and \texttt{ligo.skymap} \cite{ligoskymap}. These tools support the MOC-encoding process both of the standard HEALPix format (where all the cells are at a single order) and also the multi-resolution HEALPix format 
 \cite{singer2016}$\footnote{\url{https://emfollow.docs.ligo.org/userguide/tutorial/multiorder_skymaps.html}}$ in which different HEALPix orders are included.

\subsection{Aladin Desktop} 
Aladin Desktop is an interactive sky atlas based on Java technology. It was first created in 1999 by the CDS  and is regularly updated to be compatible with existing or emerging VO standards. The user-friendly graphical interface is shown in Figure \ref{fig:aladin_tree}. In the left-hand side of the main window is placed the Aladin data collections tree$\footnote{\url{https://www.youtube.com/watch?v=ggnJ5glhRmA}}$ designed for interactive data access and discovery. The right-hand side is called Aladin stack$\footnote{\url{https://www.youtube.com/watch?v=-p_nYH42NUo}}$ and shows all elements that have been loaded during the Aladin Desktop session. 

In the Aladin Desktop application a dedicated function is available to generate a credible region at a defined confidence level\footnote{\url{https://emfollow.docs.ligo.org/userguide/resources/aladin.html}}.
This is accomplished by selecting the menu bar the item called \textbf{Coverage $\Rightarrow$ Generate a spatial MOC based on $\Rightarrow$ The current probability skymap}.

The MOC generation window, shown in the left panel of Figure \ref{fig:window_moc}, offers three options. 

\begin{itemize}
       \item \textbf{Proba skymap}: the dropdown menu lists the image files loaded during your Aladin Desktop session. 
       To preselect a gravitational-wave sky localization in the dropdown menu, click on its name in the Aladin Stack.
       \item \textbf{Probability threshold}:  a number between 0 and 1 for the credible level
       you wish to select.
       \item \textbf{MOC resolution}: the dropdown menu lists the MOC-resolution options and the corresponding pixel resolutions.  The MOC resolution is related
       to the \textit{nside} HEALPix parameter following the equation \ref{eq:nside_order}.
\end{itemize}

Pressing the \textbf{CREATE} button, the MOC for the credible region is created and loaded in the Aladin Stack. Repetition of this process for different credible levels builds up a set of MOCs that can be selected independently from the Aladin stack.

The main Aladin window shows the credible region over a selected background-image survey for a full interactive visualisation. Among the most important interactive features is the ability to access information on the size of the region.
The area in square degrees and the percentage of the sky are shown in the top-right corner of the Aladin stack when you hover the cursor over the MOC name.
Alternatively, you can right-click the MOC in the Aladin stack and select \textbf{Properties} from the contextual menu. 

Finally, the files may be exported via the menu bar item labelled \textbf{File $\Rightarrow$ Export planes (FITS,VOTable,...)}.
The contextual window permits the selection of the relevant planes so that they may be saved in a target directory in various formats (see Subsection \ref{subsec:serialisations}).

\subsection{The \texttt{ligo.skymap} python package}
The \texttt{ligo.skymap} package  is a set of Python tools for reading, writing, generating, and visualising gravitational-wave probability maps from LIGO and Virgo. 

From \texttt{ligo.skymap}, it is possible to invoke the command-line \texttt{ligo-skymap-contour-moc}$\footnote{\url{https://lscsoft.docs.ligo.org/ligo.skymap/tool/ligo_skymap_contour_moc.html}}$.
The method returns a credible level encoded in a MOC data structure of an all-sky probability map in FITS format (see Subsection \ref{subsec:serialisations}). 

The method is based on \texttt{mocpy}. Recently, the \texttt{mocpy} version 0.10.0  builds the credibility regions with an algorithm that has various options, including 
one which produces the same results as Aladin Desktop, and others that allow for different handling of the region boundaries; see \textbf{from\_multiordermap\_fits\_file} method$\footnote{\url{https://cds-astro.github.io/mocpy/stubs/mocpy.MOC.html}}$.

\texttt{Mocpy} is a Python library allowing easy creation, 
parsing and manipulation of sky areas encoded in a MOC data structure \cite{mocpy}.  
Excellent performance in computing time is obtained thanks to the core functions written in Rust programming language$\footnote{Release 0.4.1 \url{https://www.rust-lang.org/}}$.
The library is part of the affiliated packages of the Astropy project \cite{astropy13}, \cite{astropy18}.


\subsection{MOC data visualisations}
MOC maps can be visualised in various environments and libraries. The MOC files can be viewed in the Aladin Desktop or in a dedicated Jupyter Notebook \cite{jupyter} using \texttt{ipyaladin}$\footnote{\url{https://github.com/cds-astro/ipyaladin}}$. The files may also be viewed in JavaScript applications in which Aladin Lite \cite{aladin_lite} is embedded. 
 Dedicated tutorials for plotting a general MOC map using the popular Matplotlib library \cite{matplotlib} are available in the \texttt{mocpy} repository$\footnote{\url{https://github.com/cds-astro/mocpy}}$.

The property of shading the areas inside the MOC or the property of displaying only the edges are well-suited to display/manage many overlapping planes (see Subsection \ref{subsec:updating}). 

\section{Demonstrations of the utility of MOCs in the preparation of observation campaigns}
The scheduling of observational follow-up programs requires many considerations to be taken into account. The primary consideration is the choice of the sky fields which is determined by the credible regions.  Here we discuss other constraints  that could be taken into account to develop an observing strategy: $(i)$ identifying high Galactic dust extinction areas, and $(ii)$ locating  regions in which reference images are already available.

\subsection{Application example of GW190425}
We  apply the method to identify observable sky zones in a MOC map in the context of the low-latency gravitational-wave alert of GW190425 \cite{gw190425}. 
The event represents the discovery of a second binary neutron star merger after GW170817 \cite{gw170817}. The GW190425 signal has been observed on \texttt{2019 April 25, 08:18:05 UTC},  during the third observing run (O3) of the LIGO–Virgo network. The network consists of two Advanced LIGO interferometers 
in Hanford, Washington, USA (LHO) and Livingston, Louisiana, USA (LLO) and the Advanced Virgo
interferometer in Cascina, Italy. At the time of GW190425, LHO was temporarily offline with only LLO and Virgo taking data. 

The initial sky map, generated by BAYESTAR algorithm \cite{singer2016}, has a 90\% credible region of 10,183 deg$^2$ \cite{gcn24168}. 
It is displayed on the top of Figure \ref{fig:skymap_operations_1}. Up to the present day, no confirmed electromagnetic or neutrino event has been identified in association
with this gravitational-wave event \cite{gw190425}. 

\begin{figure}
\center
	\includegraphics[width=0.8\textwidth,]{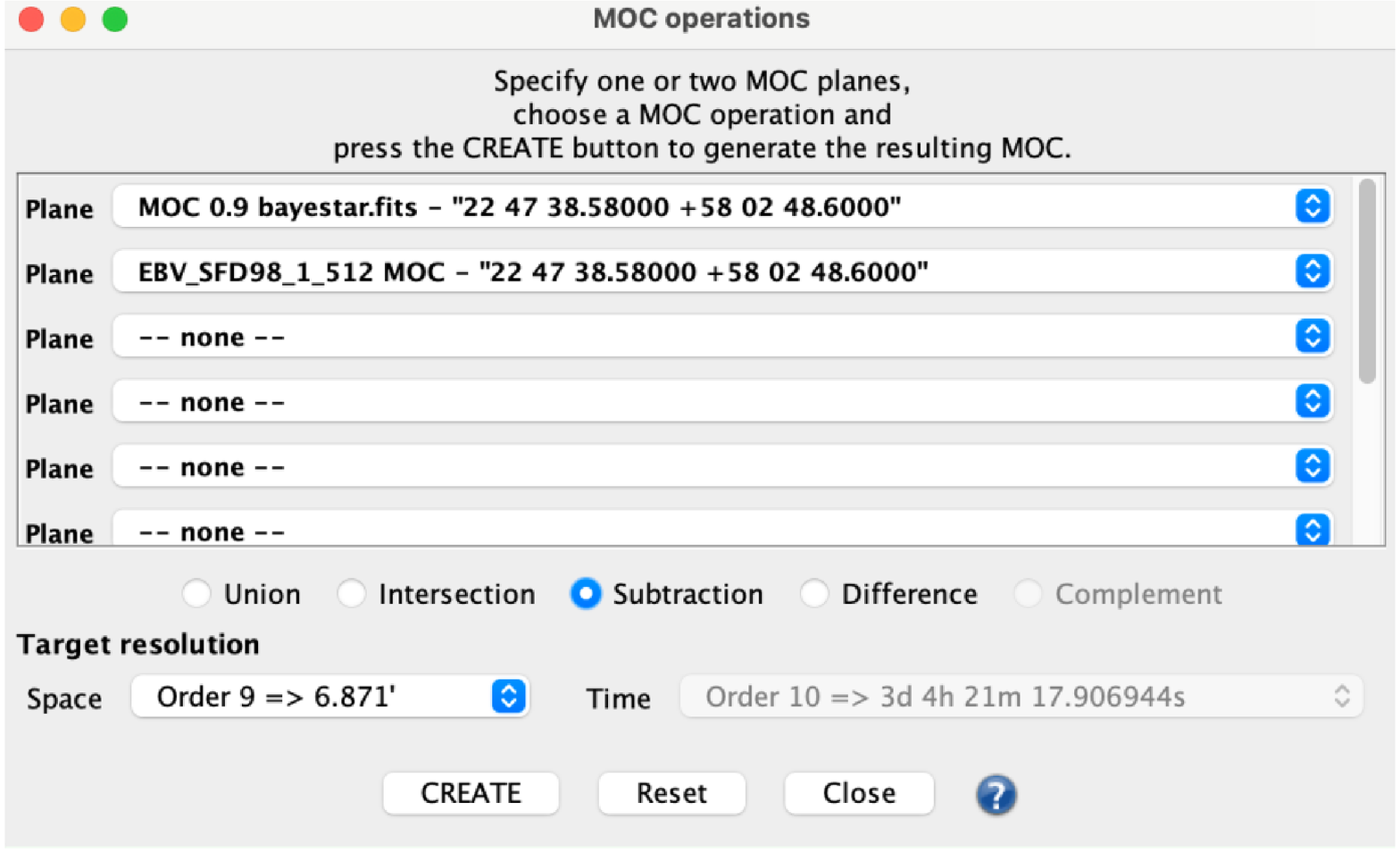}
	\caption{MOC operations. The window provides a set of MOC functions: union, intersection, subtraction and difference for manipulating spatial and/or temporal coverages  by selecting the corresponding radio button. The planes are selectable from the dropdown menus. In \textbf{Target resolution}, different MOC orders for space and/or time resolutions can be independently chosen.}
	\label{fig:logical_operation}
\end{figure}

\subsection{Identifying high dust extinction areas}
\label{dust_remove}
Searches for transient objects generally need to take into account factors that affect the detectability of sources, such as the galactic dust extinction and the stellar density \cite{grawita}, \cite{Andreoni_2020}. The algorithm developed by various groups to construct electromagnetic follow-up strategies (\cite{salafia17}  \cite{coughlin18}, \cite{decam})  emphasise the impact of dust extinction and stellar density as variables on the detectability of electromagnetic transients associated with gravitational-wave sources. The example described in the present paper is admittedly oversimplified but at the same time provides step-by-step guidelines to customise new results and proper working processes.
 
For the purpose of the general application, 
we  download the  all-sky Galactic reddening map from \citep{Schlegel} in HEALPix sky pixelization scheme\footnote{\url{http://cade.irap.omp.eu/dokuwiki/doku.php?id=galactic_reddening}} from the Analysis Center for Extended Data (CADE)\footnote{\url{http://cade.irap.omp.eu/dokuwiki/doku.php?id=start}}. The ingestion tool adopted by CADE is based on the drizzling library\footnote{\url{http://cade.irap.omp.eu/dokuwiki/doku.php?id=software}}, and applies an algorithm where the surface of pixel intersection is computed as described in \citep{paradis2020}. The service delivers VO-compatible data, accessible through the Aladin software.

With the all-sky map of Galactic reddening loaded into Aladin Desktop, the MOC  is computed from a pixel range as follows: select \textbf{Coverage $\Rightarrow$ Generate a spatial MOC based on $\Rightarrow$ The current  image, HEALPix map or HIPS}.
The MOC generation window (shown in 
the right panel of Figure \ref{fig:window_moc}) offers three options. 

\begin{itemize}
       \item \textbf{Plane}: the  dropdown  menu  lists  the  image  files  loaded in the Aladin  Desktop  session. To preselect a plan, click on its name in the Aladin Stack.
       \item \textbf{Pixel range}: a range of pixel values.
       \item \textbf{MOC resolution}: the dropdown menu lists the MOC-resolution options and the corresponding pixel resolutions.  The MOC resolution is related to the \textit{nside} HEALPix parameter following the equation \ref{eq:nside_order}.
\end{itemize}

Pressing the \textbf{CREATE} button, the resulting MOC is generated and loaded in the Aladin stack.

In this case, we have chosen the range of pixel values from 0.5 up to the highest value, producing an area of 5,472 deg$^{2}$. The resulting MOC from the extinction map, overlapping the original gravitational wave sky localization, is shown in the middle panel of Figure \ref{fig:skymap_operations_1}.
 
The operation between the sky areas is performed by selecting \textbf{Logical operation} from \textbf{Coverage} in the main menu. The window is shown in Figure \ref{fig:logical_operation}. 
The \textbf{MOC operations} window provides a set of MOC functions: union, intersection, subtraction and difference for manipulating spatial and/or temporal coverages by selecting the corresponding radio button. The planes are selectable from the dropdown menus. In \textbf{Target resolution}, you can independently choose different MOC orders for space and/or time resolutions. (For more details in the two-dimensional interleaving approach see Section 3.3 in IVOA  Proposed Recommendation  \cite{fernique21}.)

The resulting coverage map, shown at the  bottom panel of Figure~\ref{fig:skymap_operations_1} depicts the initial sky localization of GW190425 in which certain Galactic reddening values have been subtracted. The operation is computed with a space MOC order = 9.  (As no temporal information is stored in those MOCs; the time resolution dropdown menu appears disabled as shown in Figure \ref{fig:logical_operation}.)

\subsection{Searching for reference images}
\begin{figure}
   \center
	\includegraphics[width=1.0\linewidth]{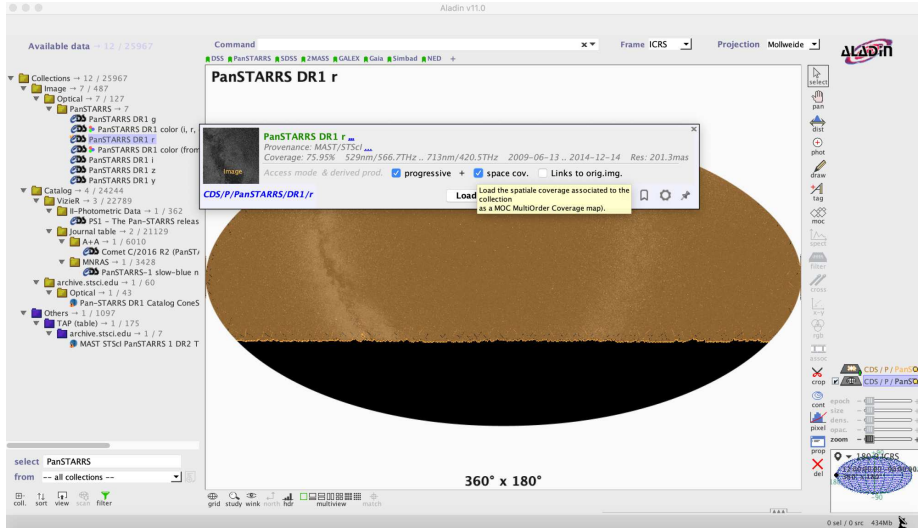}
	\caption{Aladin Graphical User Interface. Left-hand window: exploring the Aladin data collections   tree. When a catalogue item has been selected, 
	 we can also require to download the spatial coverage
	 associated to the survey as a MOC map checking the box \textbf{space cov.}.  The data provenance is reported in the same contextual window: \texttt{Provenance: MAST/STScl}. By clicking on \textbf{...}, visualised in the same line,
	 additional information is linked. Left-hand window: Aladin stack. It stacks the loaded planes in an Aladin section.}
	\label{fig:aladin_tree}
\end{figure}

To search for convenient reference images, we explore the public surveys from the Aladin data 
collections tree$\footnote{\url{https://www.youtube.com/watch?v=IG_6Eh9EKKk}}$. When one or more image resources  have been chosen, 
the spatial coverage related to the resource is loaded 
as a MOC map by checking the box \textbf {space cov.} (see Figure \ref{fig:aladin_tree}).  This functionality permits further operations to be done  
with the new MOC map. As a first step,  we choose  the PanSTARRS DR1$\footnote{\url{https://panstarrs.stsci.edu/}}$ coverage and then we intersect it with 
the BAYESTAR gravitational-wave sky localization  of GW190425 \textit{pre-processed}  as discussed in the previous Section \ref{dust_remove}.
 
The MOC map of the PanSTARRS DR1 dataset has been generated by CDS as part of the ingestion of the PanSTARRS data into the CDS HiPS node. The MOC was generated using Aladin/HipsGen v10.138.

\begin{figure}
	\includegraphics[width=\linewidth]{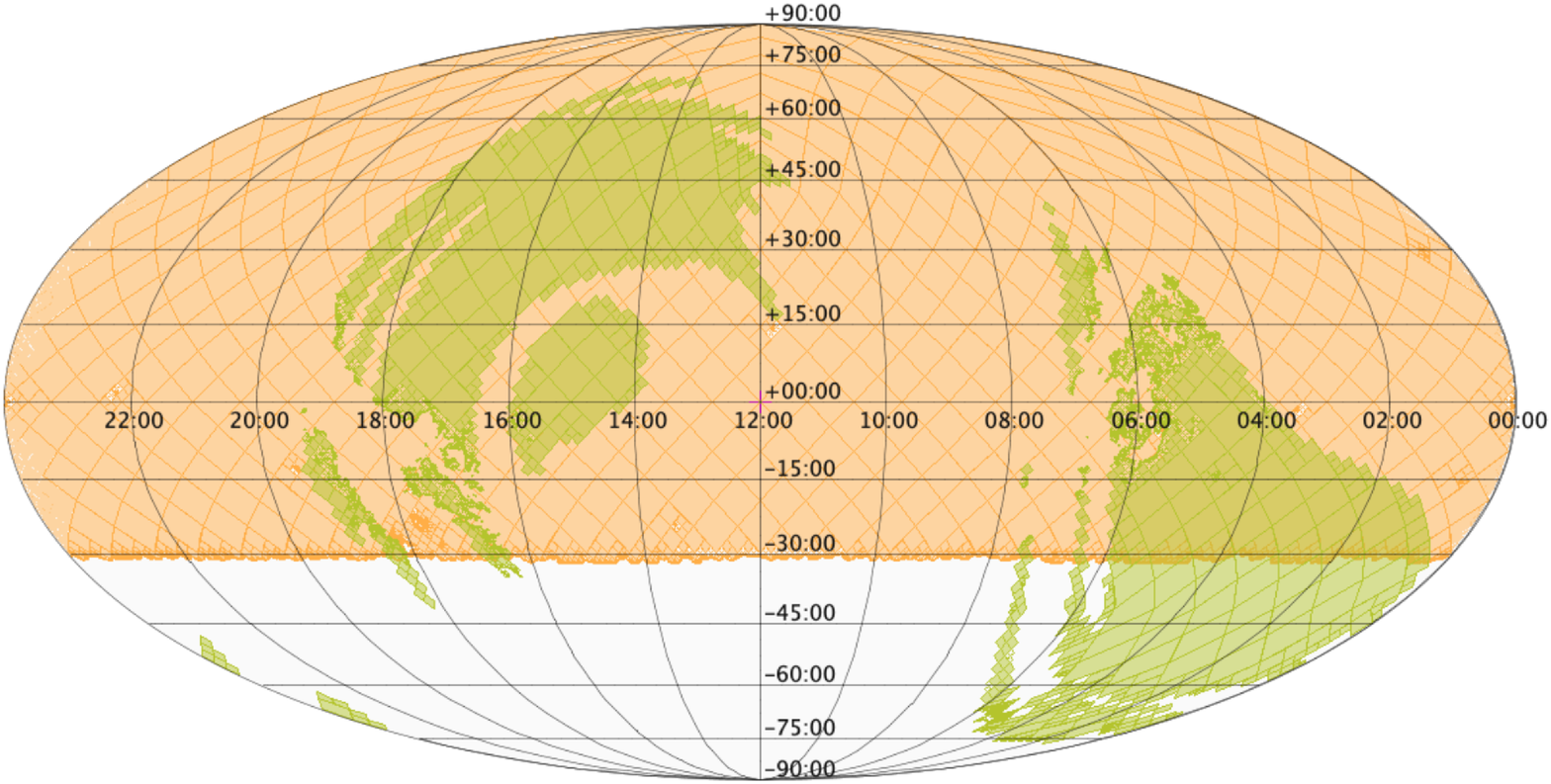}
	\includegraphics[width=\linewidth]{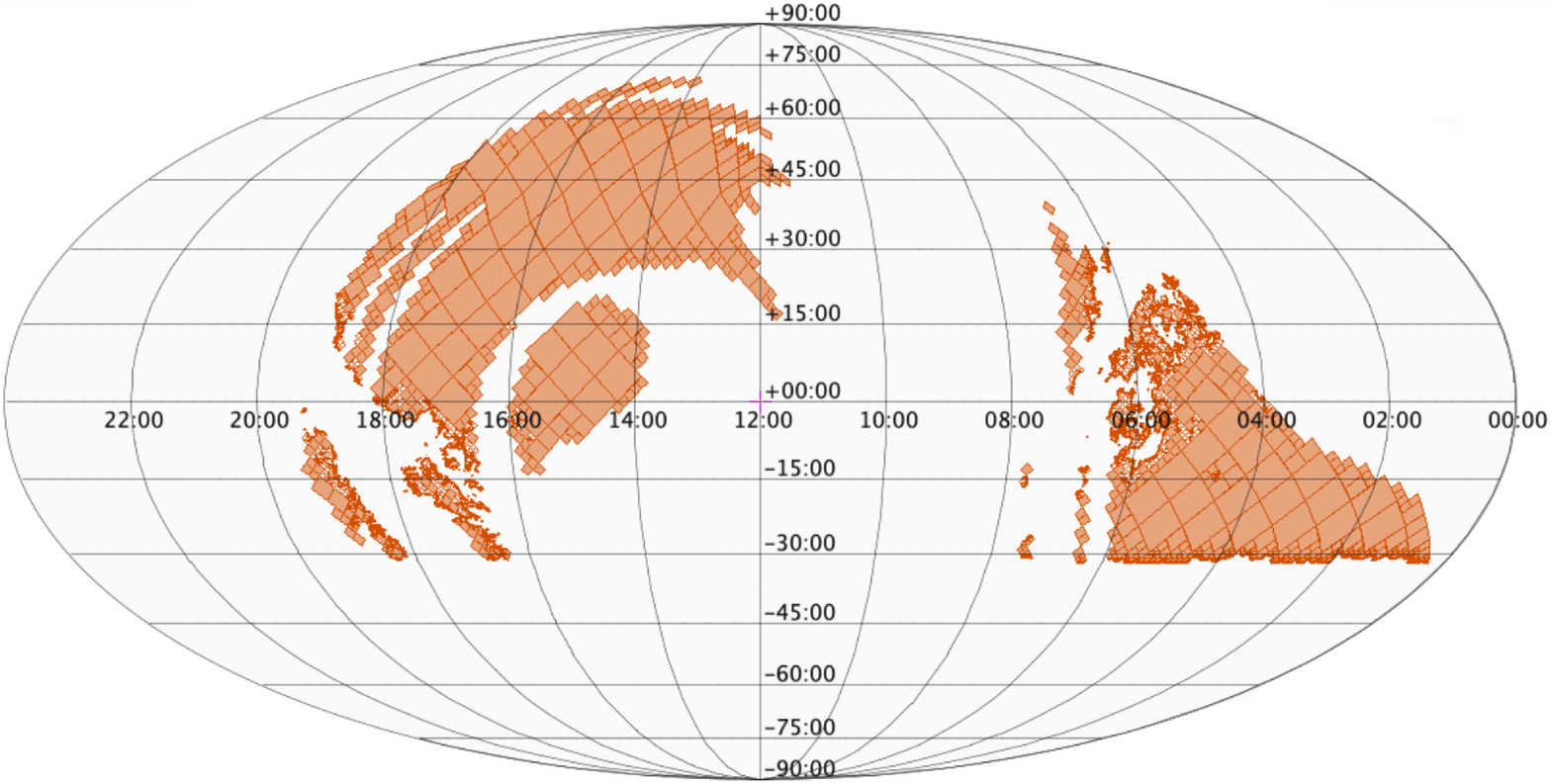}
	\caption{\textit{Top}. In orange the PanSTARRS DR1 coverage of  31,330 deg$^2$ overlapping the  
	gravitational-wave sky localization of GW190425 in which  
	dust absorption fields have    been  removed. \textit{Bottom.} The resulting intersection area of 6,777 deg$^2$ (MOC order = 11).}
	\label{fig:skymap_operations_2}
\end{figure}

Figure~\ref{fig:skymap_operations_2}
depicts the MOCs to be used for the operational planning. 
The top panel shows the PanSTARRS DR1 coverage (in orange)  overlapping the  
initial gravitational-wave sky localization of GW190425 in which  dust absorption fields have been  removed.  
The bottom panel  displays the intersection area between the two sky coverages.
The resulting sky map has been produced by using the  \textbf{Logical operation} window in Figure \ref{fig:logical_operation}. The options available in this window are described in Section~\ref{dust_remove}.
The operation is computed with a space MOC order = 11.  No temporal information is stored in that MOCs; the time resolution dropdown menu  in Figure \ref{fig:logical_operation} will be appear disabled.
 
This approach provides a very efficient method to identify sky areas in which reference images are available for searching for potential electromagnetic transients associated with a gravitational-wave source. In Section \ref{subsec:retrieving},
we will describe how to gather images when the MOC operations have been concluded according to strategies adopted in an electromagnetic follow-up, and the visibility from an observational site is computed.

\section{Determining multi-observability within a MOC coverage}
Here we describe the essential steps to delimit portions of a MOC coverage that are observable from a specific astronomical site within a fixed time interval and a defined range of airmass values. For this example, we start with the MOC that was described in the previous section. The complete code is reported in a public GitHub repository in which a jupyter notebook is provided$\footnote{\url{https://l.infn.it/tuto-moc-mma}}$.

\subsection{Flattening a MOC map at a fixed order} 
A MOC object is a collection of  cells, originated from the HEALPix tessellation, and stored at different $orders$. In fact, the list of cells  is  compressed by replacing any
four consecutive cells at $order$ $k$ by their parent cells from $order$
$k$ - 1. This operation can be done recursively down to $k$ = 0. 
The deepest level is set by the resolution required for describing the data set \citep{moc1.1}. 

For some purposes, it is more convenient to operate on a "flattened" MOC where all of the cells are at the same order. To calculate the airmass in sky regions with equal areas, we flatten the spatial MOC map to a predefined $order$.
MOC uses the NESTED numbering scheme and, thus, each spatial MOC cell can be
stored as one single  interval ranging from $[npix \times 4^{(max\_order - order)}]$ to 
$[(npix+1) \times 4^{(max\_order - order)}-1] $.  The maximum order of a spatial MOC cell is
$max\_order = 29$.

In this example, we use the  MOC  input region serialised in a JSON format.  Subsequently, the JSON file is read as an ordinary Python dictionary 
with the \textit{keys} and \textit{values} representing the \textit{orders} and  the \textit{pixel indexes}, respectively. 

\subsection{From MOC cells to sky coordinates}
When the  entire collection of cells in a MOC map is flattened to the same order,  we invoke from \texttt{cds-healpix-python} module 
the \textbf{healpix\_to\_skycoord} method \cite{cds_healpix}. This method converts HEALPix indices to celestial coordinates.
The input is a 1-D array of HEALPix indices and the resulting celestial coordinates are returned in a \textbf{SkyCoord} high-level object as defined in astropy \cite{astropy18}.

\subsection{Computing the visibility of MOC regions from observatory sites}
The Python module astroplan \cite{astroplan2018} is used to compute the visibility of a MOC from different observatories by specifying the location and the airmass constraints.

We consider a multi-messenger team composed of three astronomical research observatories: 1) Haleakala Observatories in Hawaii, USA 
2) Paranal Observatory in Chile 3) Siding Spring Observatory (SSO) in Australia. 
With \texttt{astroplan} we define  three \textit{Observer} classes. Observer is an astroplan container class for information about 
an observer’s site$\footnote{\url{https://astroplan.readthedocs.io/en/latest/api/astroplan.Observer.html}}$.  

Using the methods  \textbf{at\_night}, we verify if it is night at the time of the GW190425 at  Haleakala Observatories, otherwise, we 
ask for the nearest twilight astronomical evening with \textbf{twilight\_evening\_astronomical} method. 
 
Then we calculate the airmass in steps of two hours for each  \textbf{SkyCoord} as described in  \textbf{3.1.4} of astroplan  documentation$\footnote{Release 0.4.1 \url{https://astroplan.readthedocs.io/_/downloads/en/v4.1/pdf/}}$.

\begin{figure}
\center
	\includegraphics[width=0.32\textwidth, ]{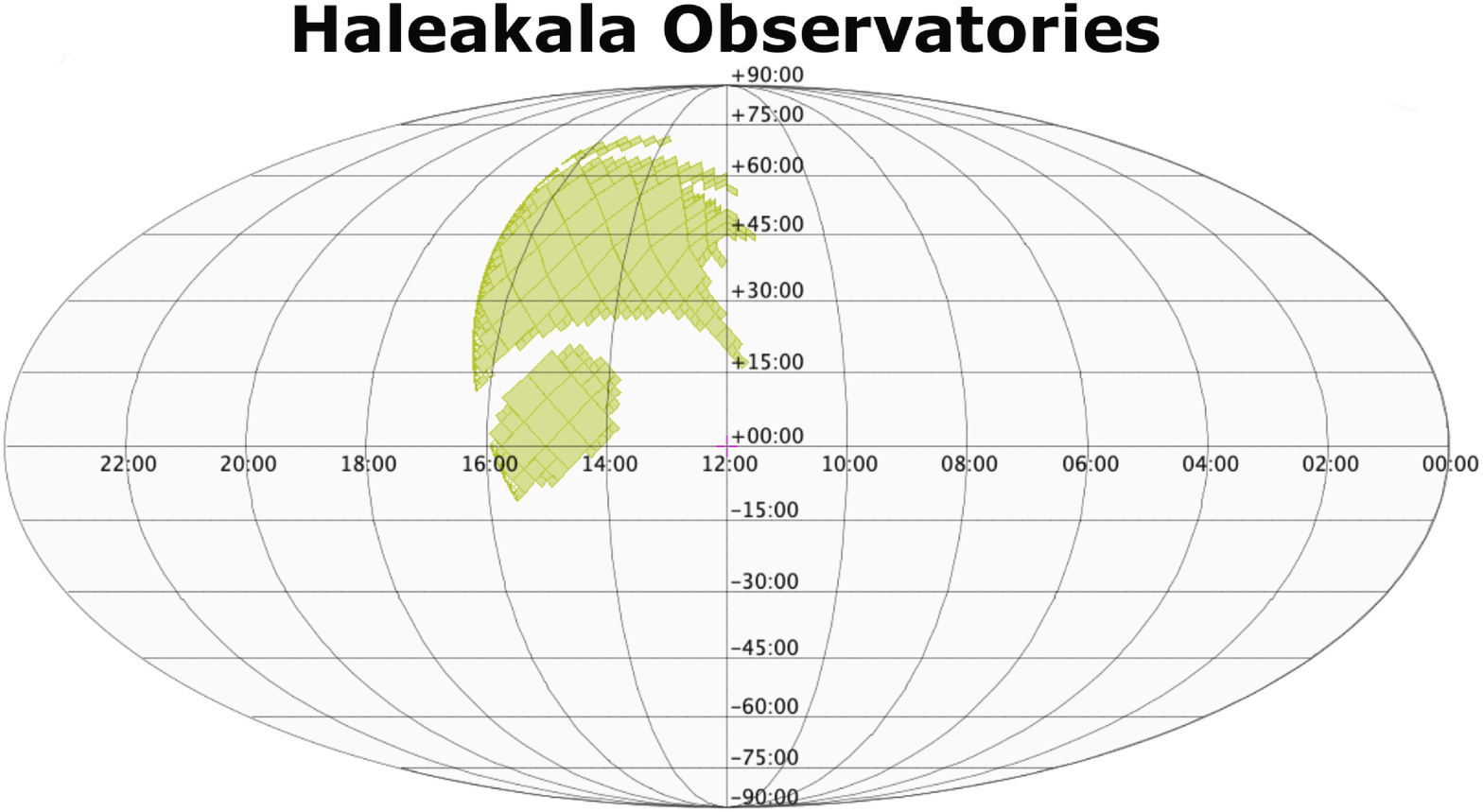}
	\includegraphics[width=0.32\textwidth,]{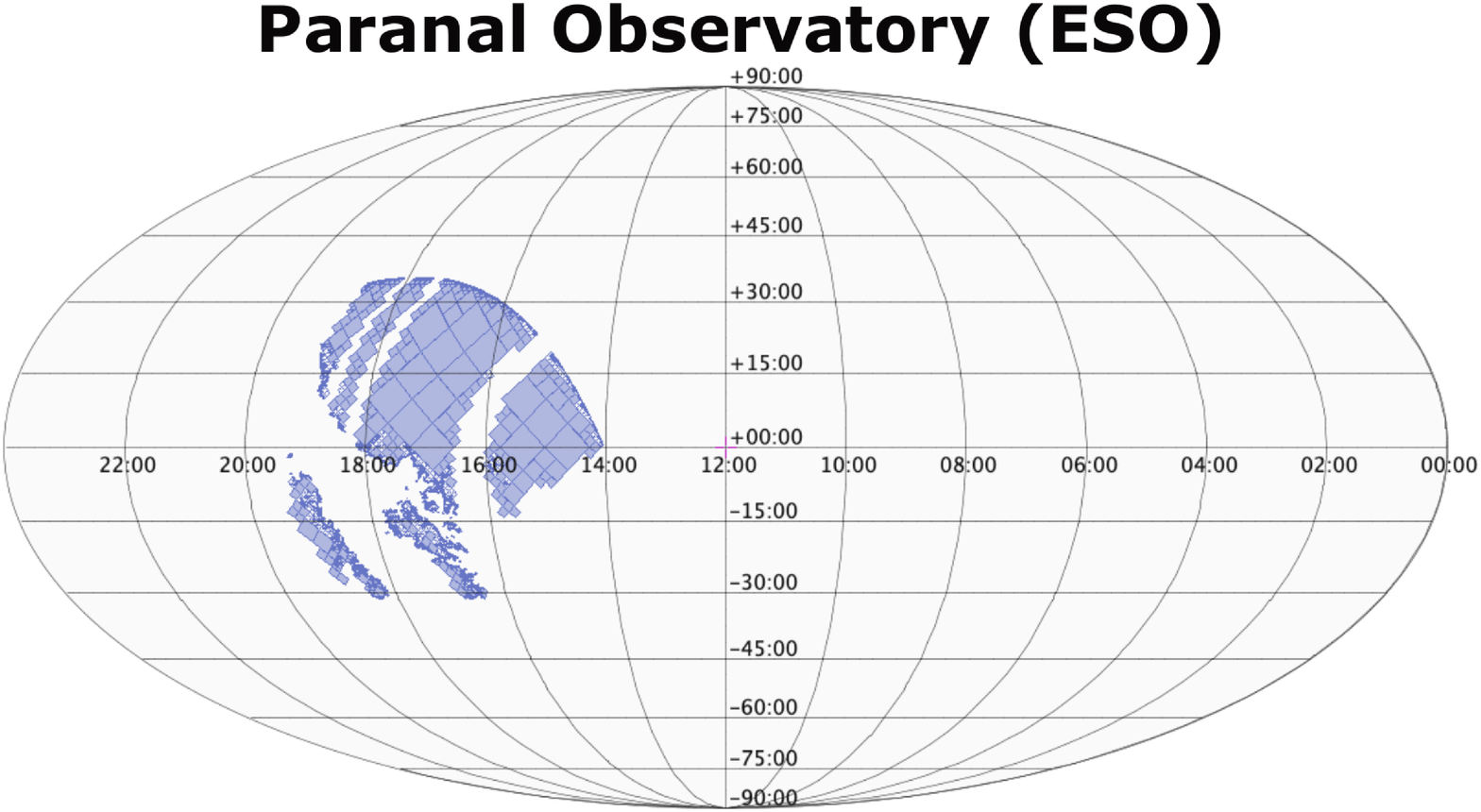}
	\includegraphics[width=0.32\textwidth,]{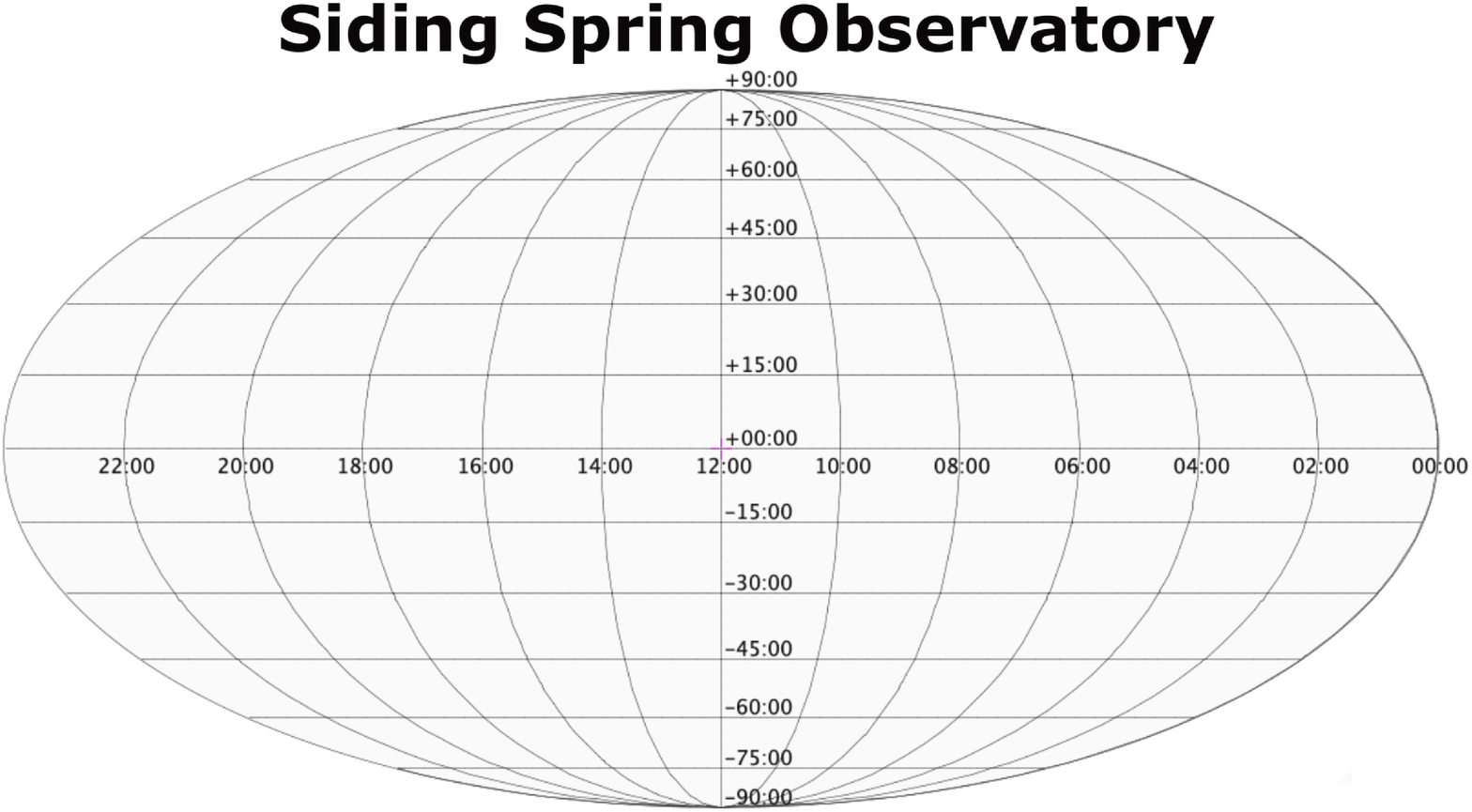}
	\includegraphics[width=0.32\textwidth, ]{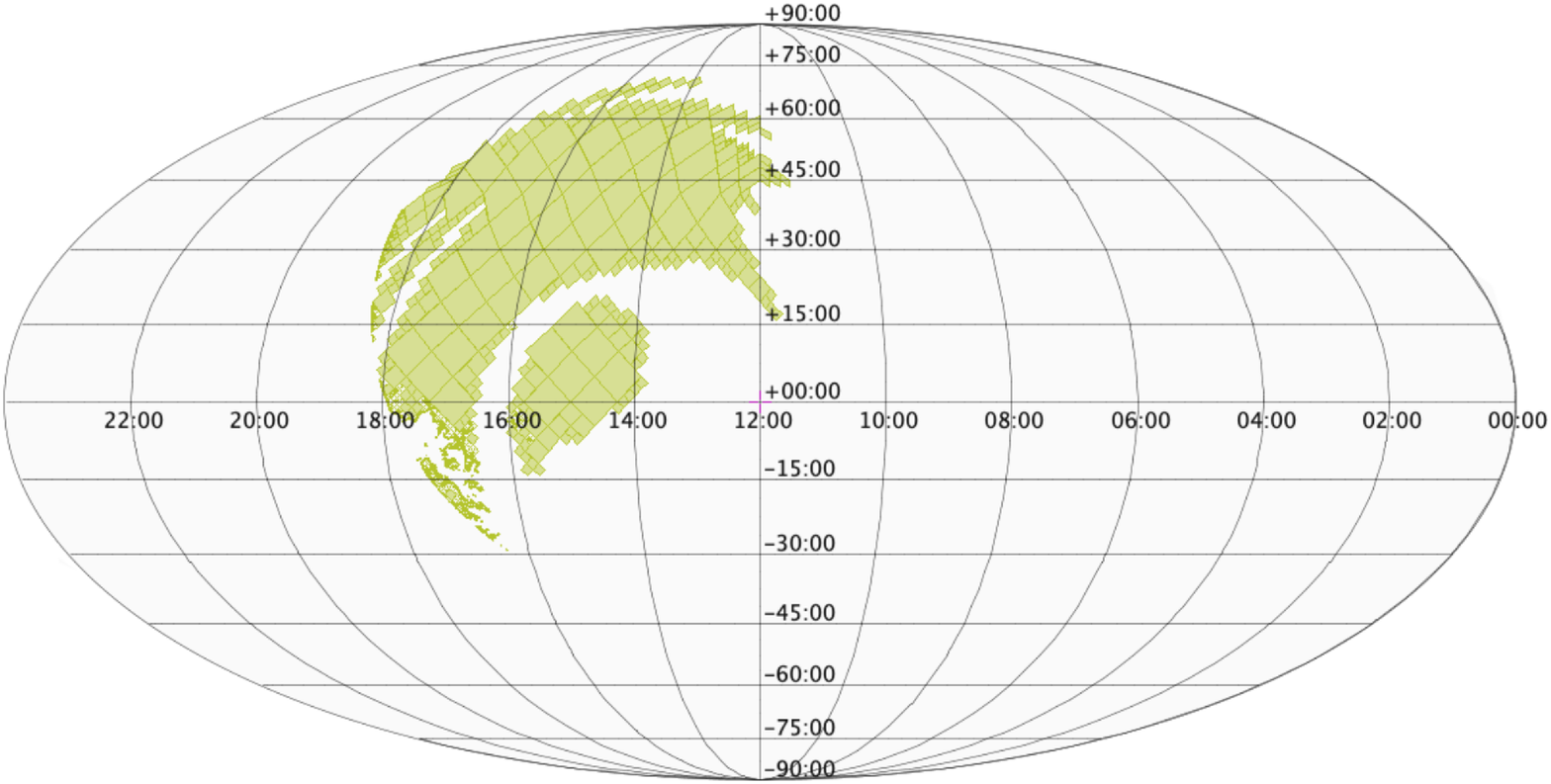}
	\includegraphics[width=0.32\textwidth,]{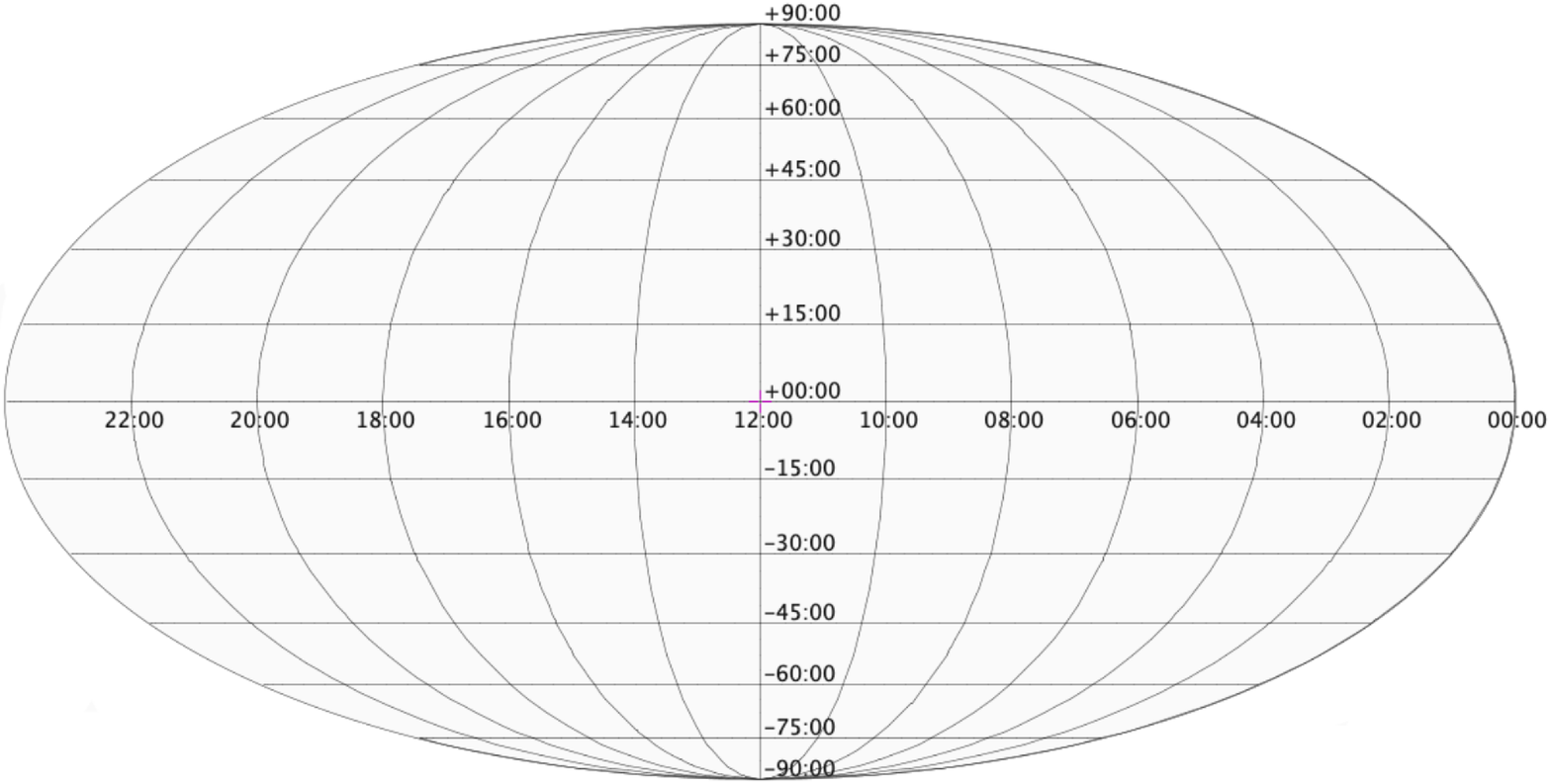}
	\includegraphics[width=0.32\textwidth,]{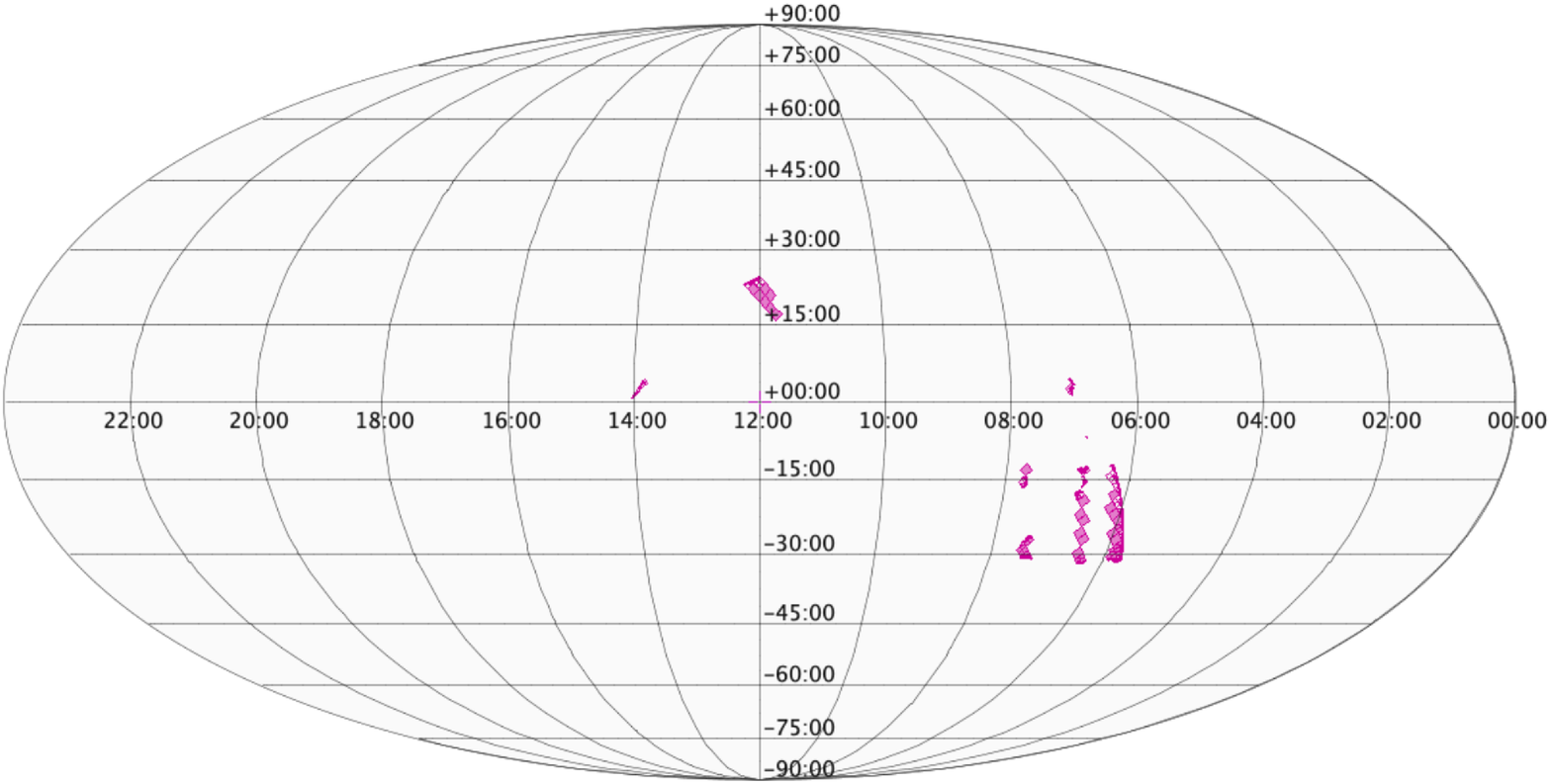}
		\includegraphics[width=0.32\textwidth, ]{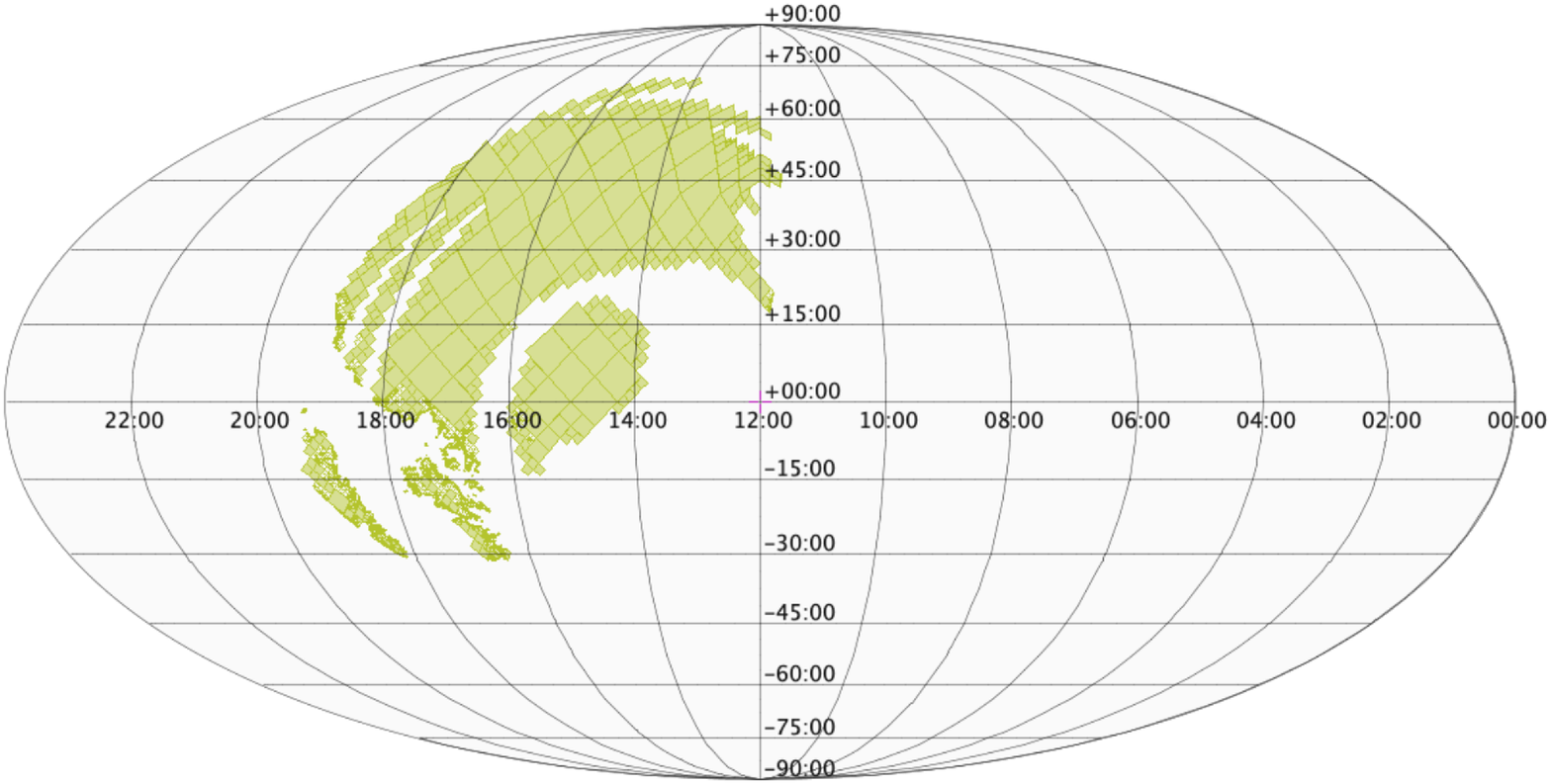}
	\includegraphics[width=0.32\textwidth,]{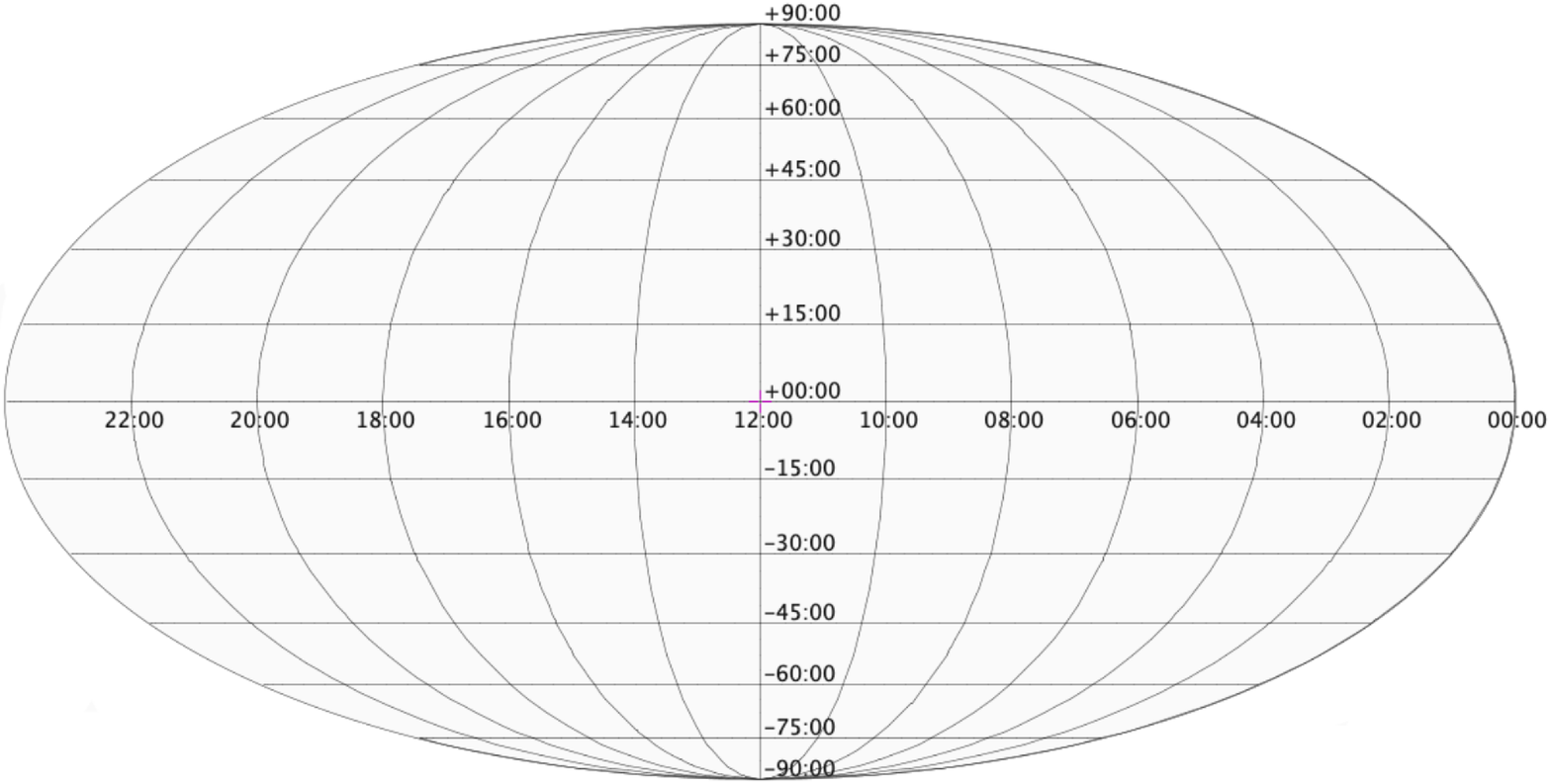}
	\includegraphics[width=0.32\textwidth,]{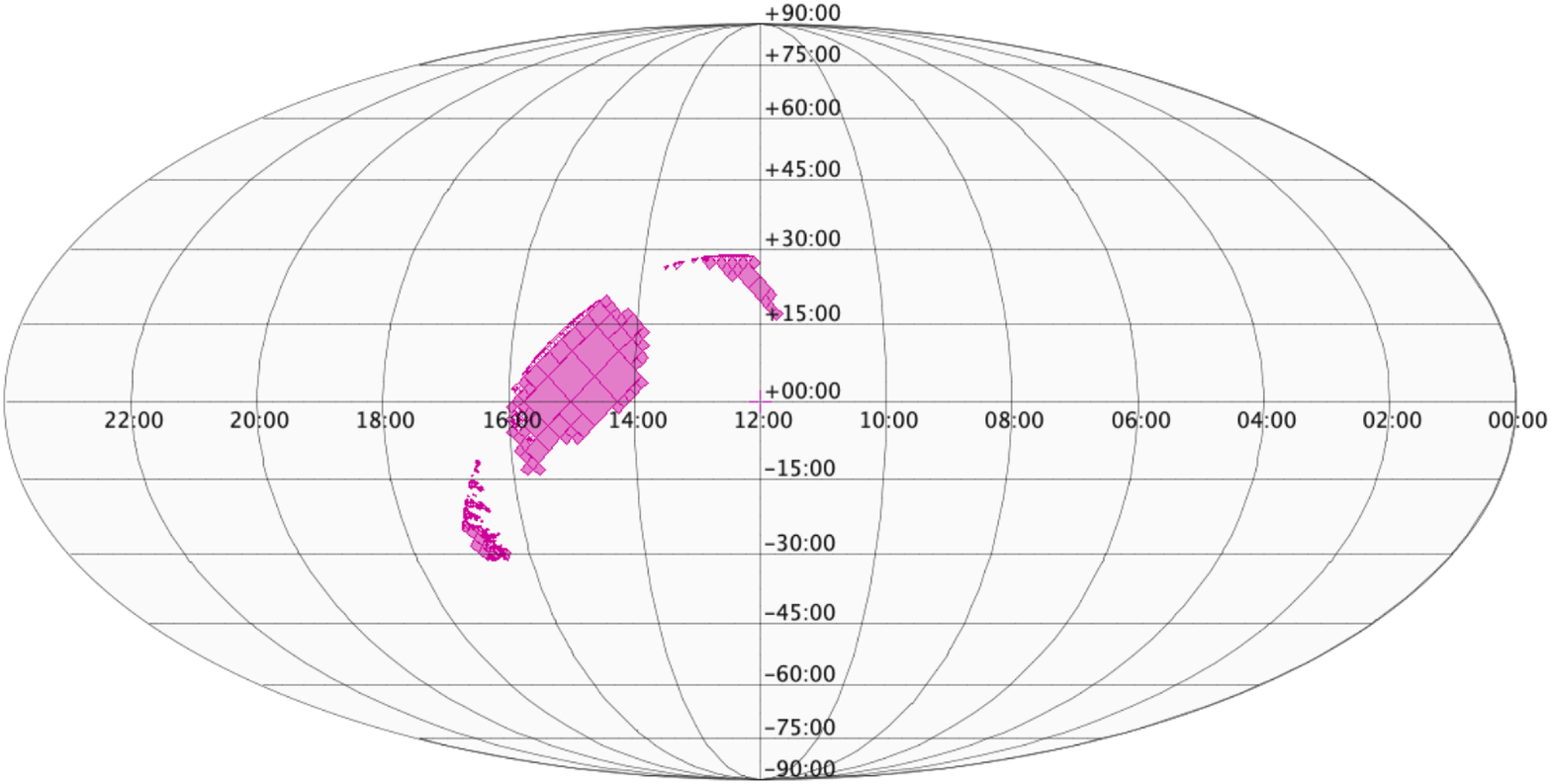}
		\includegraphics[width=0.32\textwidth, ]{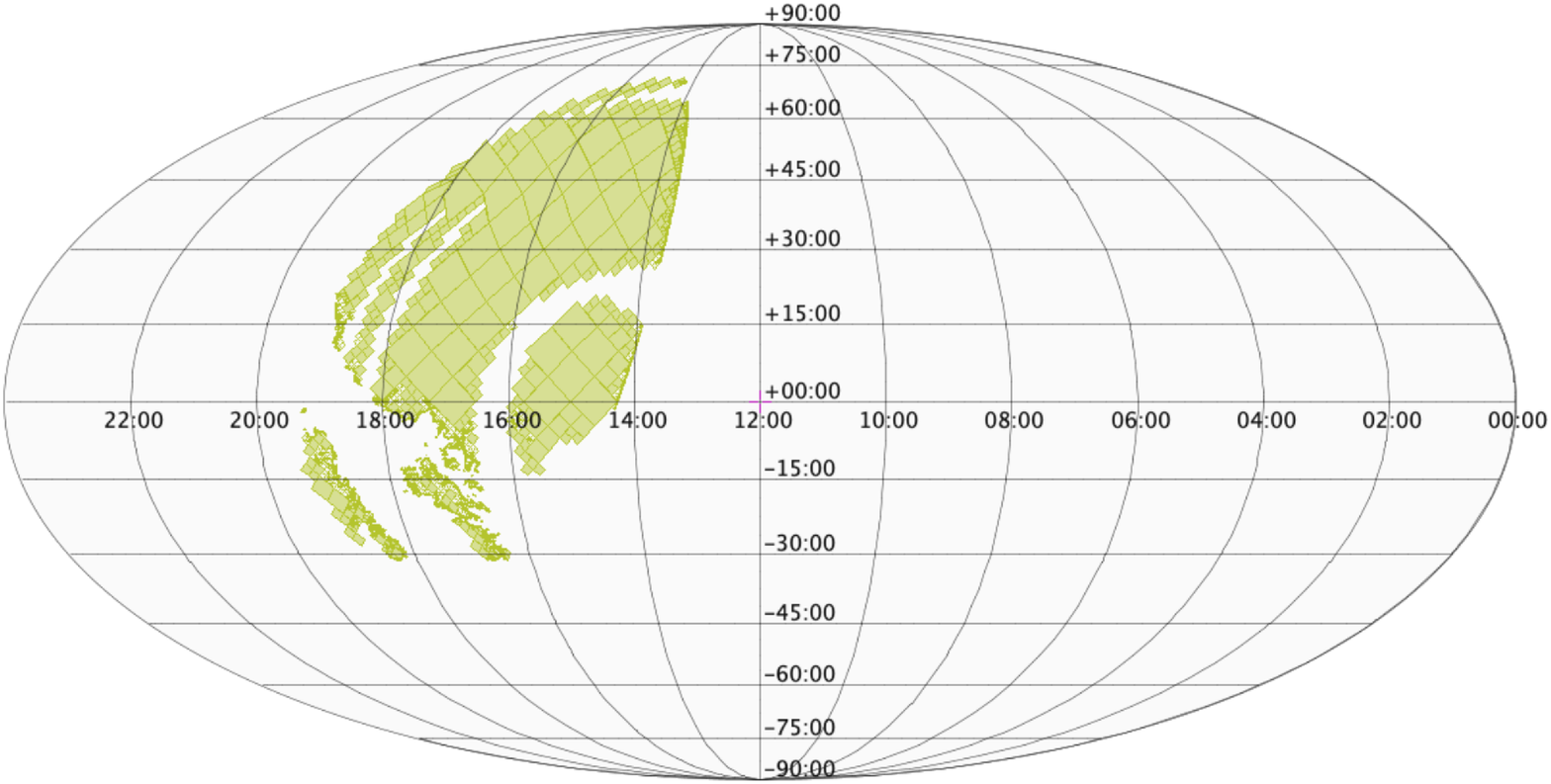}
	\includegraphics[width=0.32\textwidth,]{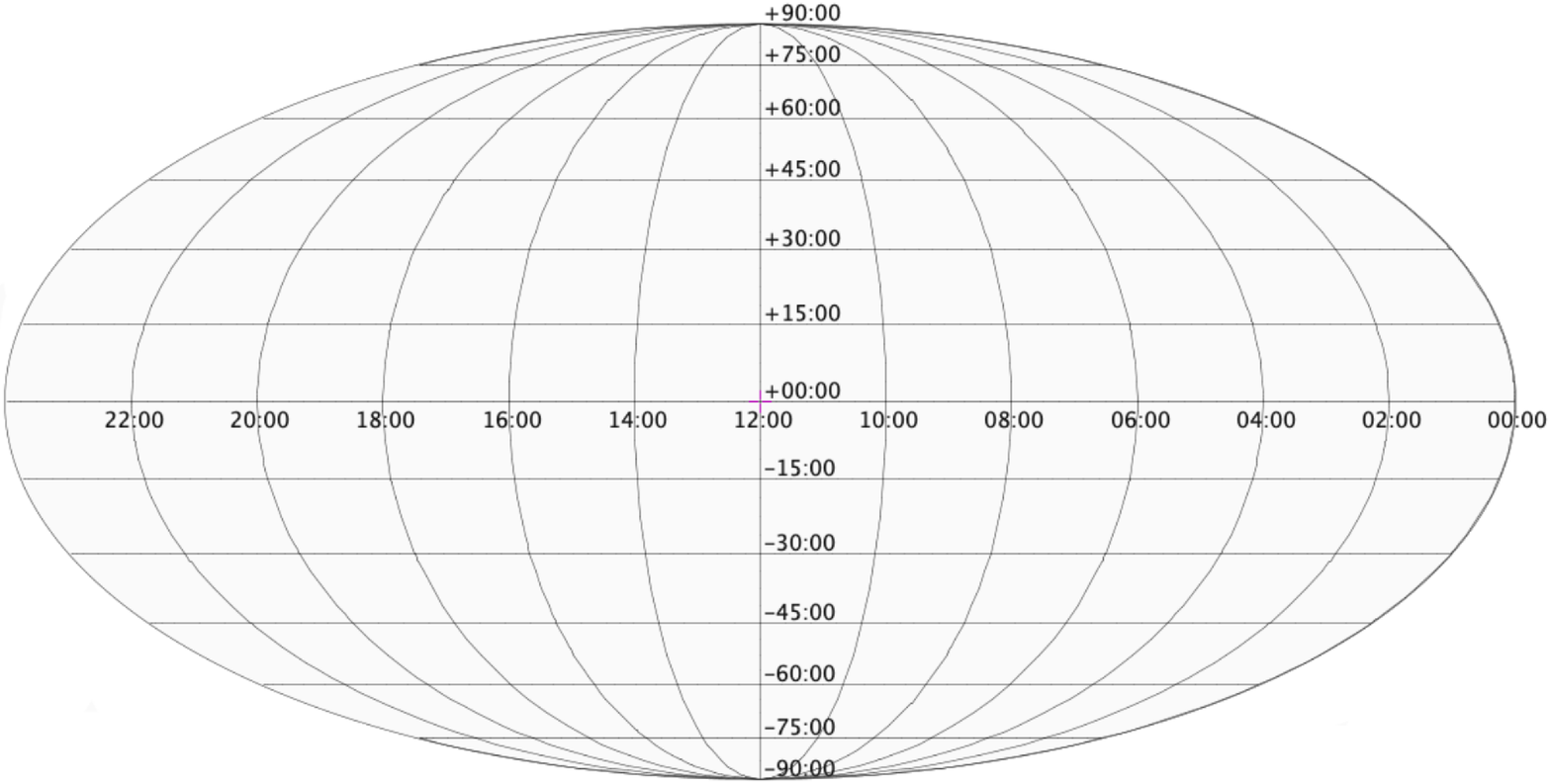}
	\includegraphics[width=0.32\textwidth,]{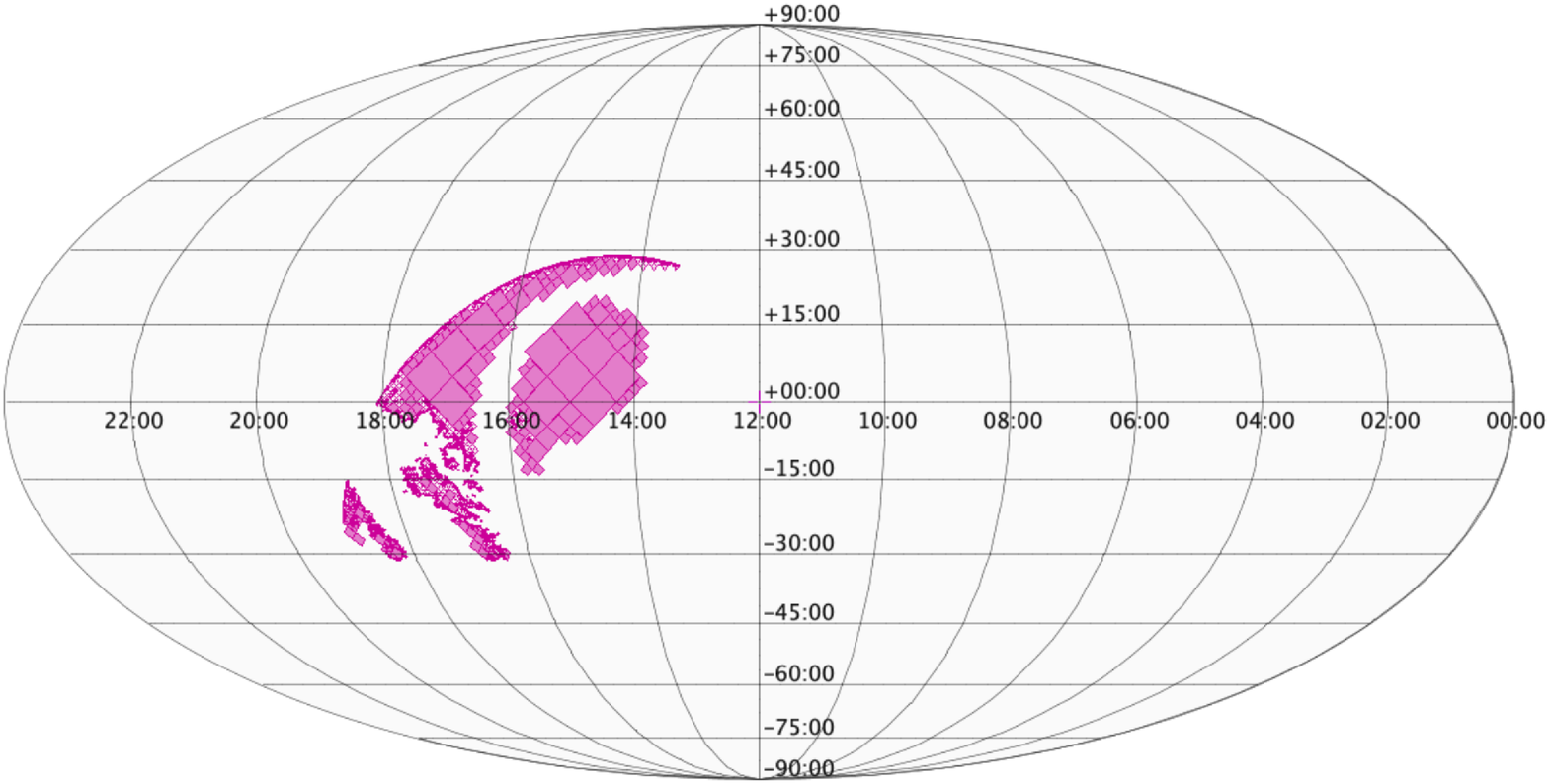}	
	\caption{MOC observabilities of the gravitational-wave sky localization of GW190425. The original sky map was previously processed taking into account  the all-sky Galactic reddening map from \citep{Schlegel} and overlapping the PanSTARRS DR1 survey as reference images. The visibility refers to three astronomical observatories: 
	Haleakala Observatories in Hawaii (USA), Paranal Observatory in Chile and Siding Spring Observatory (SSO) in Australia. The time interval is defined from 
	\texttt{08:18:05 UTC} to
	\texttt{14:18:05 UTC},  from top to bottom, in two hour steps with airmass  $1 \leq X \leq 2$. }
	\label{moc_visibility} 
\end{figure}

We create a three-column astropy table that merges  \textbf{SkyCoord} and the related airmass measurements.  The resulting table is filtered to specify cases where airmass values are in the range from 1 to 2.  
Finally, using \textbf{from\_skycoords} method in \texttt{mocpy}, we recreate a new MOC which we refer to as the visibility MOC. We repeat the same steps for Paranal and SSO observatories fixing the same starting time as in the Haleakala site. Figure \ref{moc_visibility} shows the resulting MOC visibilities from \texttt{08:18:05 UTC} to
\texttt{14:18:05 UTC}, from top to bottom, in two hour steps for each astronomical observatory.
	
Table \ref{table:areas} summarises the sky areas of the MOCs  and Table \ref{table:intersection} lists their overlaps in square degrees. As indicated by the blank values in the second column of Table~\ref{table:intersection} there are no simultaneous overlaps between all three observatories at any time, but columns 3 and 4 show significant intersections between Haleakala and Paranal, and Haleakala and SSO.
The generations of these simple tables allow organising strategies among several observers. Continuing the illustration of this example, to optimise the observations planning in such a case, the intersection areas can be totally subtracted from Haleakala schedule and redistributed between  Paranal and SSO observatories. Viceversa, for transient classifications, the same sky regions can be monitored at different wavelengths or instruments (\cite{magic}, \cite{vista}).

\subsection{Retrieving the reference images}
\label{subsec:retrieving}

The images available via the Aladin data collections tree are HiPS (Hierarchical Progressive Survey, \cite{hips}) data sets that comply with the IVOA HiPS standard \citep{hips_ivoa}. The HiPS network is made up of 20 nodes hosted at different sites, and a listing of all of the data sets published in the network (more than 900 data sets) is available\footnote{\url{https://aladin.u-strasbg.fr/hips/list}}. The provenance of each HiPS data set is included in the standardised metadata in the properties file of each data set. Cut-out images, in the FITS (Flexible Image Transport System) format, may be extracted from HiPS data sets using the online hips2fits service \citep{hips2fits}\footnote{\url{http://alasky.u-strasbg.fr/hips-image-services/hips2fits}}. 
This provides a very efficient and flexible way to generate cut-outs from many surveys, which can be useful for many purposes albeit with a number of caveats. Note that the extracted FITS files do differ from the original data because they have been converted to HiPS which involves re-sampling onto the HEALPix grid, and then extracted with user-defined sampling into a FITS cut-out by hips2fits. These transformations would need to be taken into consideration when making scientific measurement from these FITS files.

\begin{figure}
\label{fig:tiling_retrieving}
\center
	\includegraphics[width=0.8\textwidth, ]{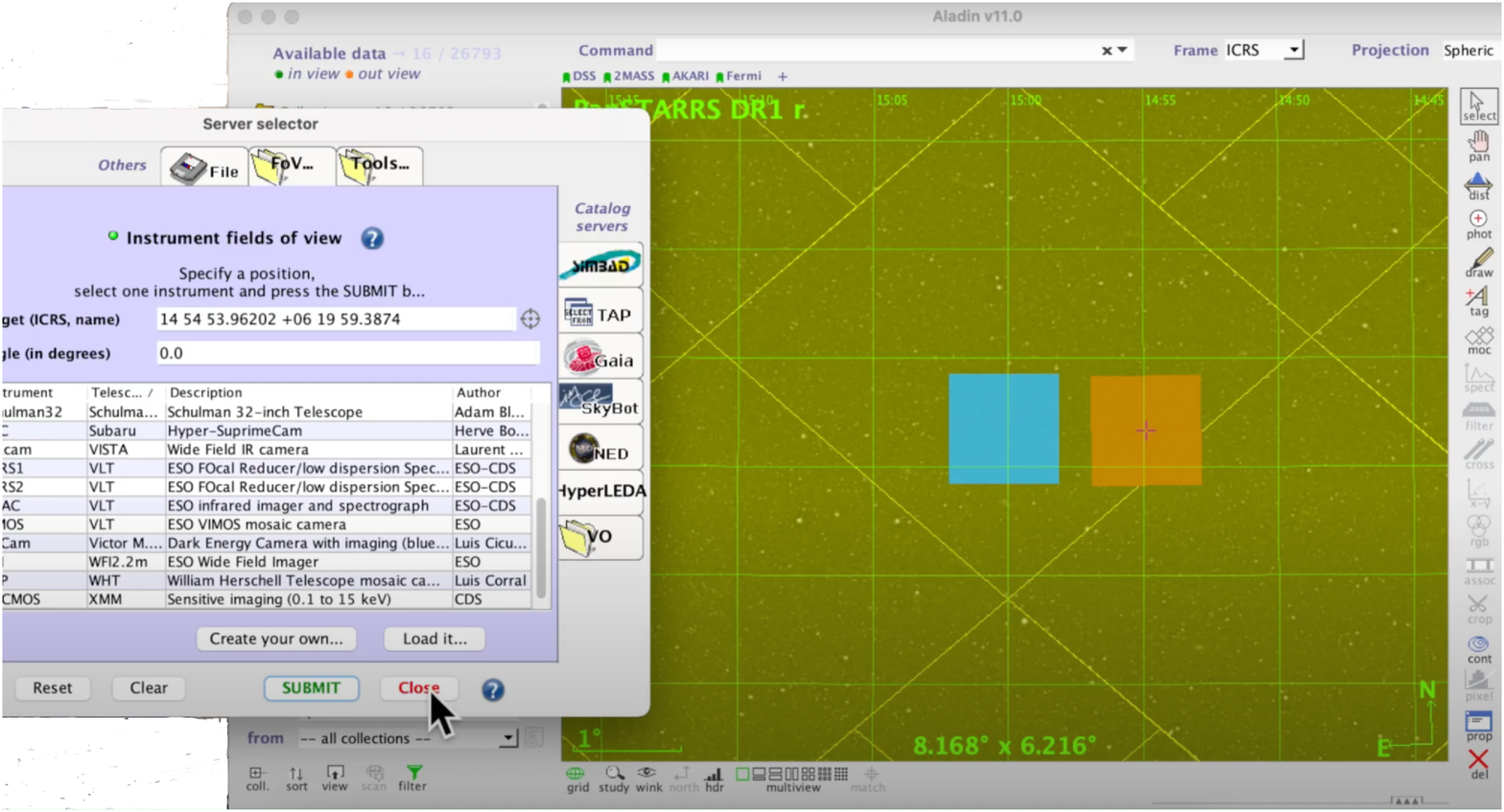}
	\includegraphics[width=0.8\textwidth,]{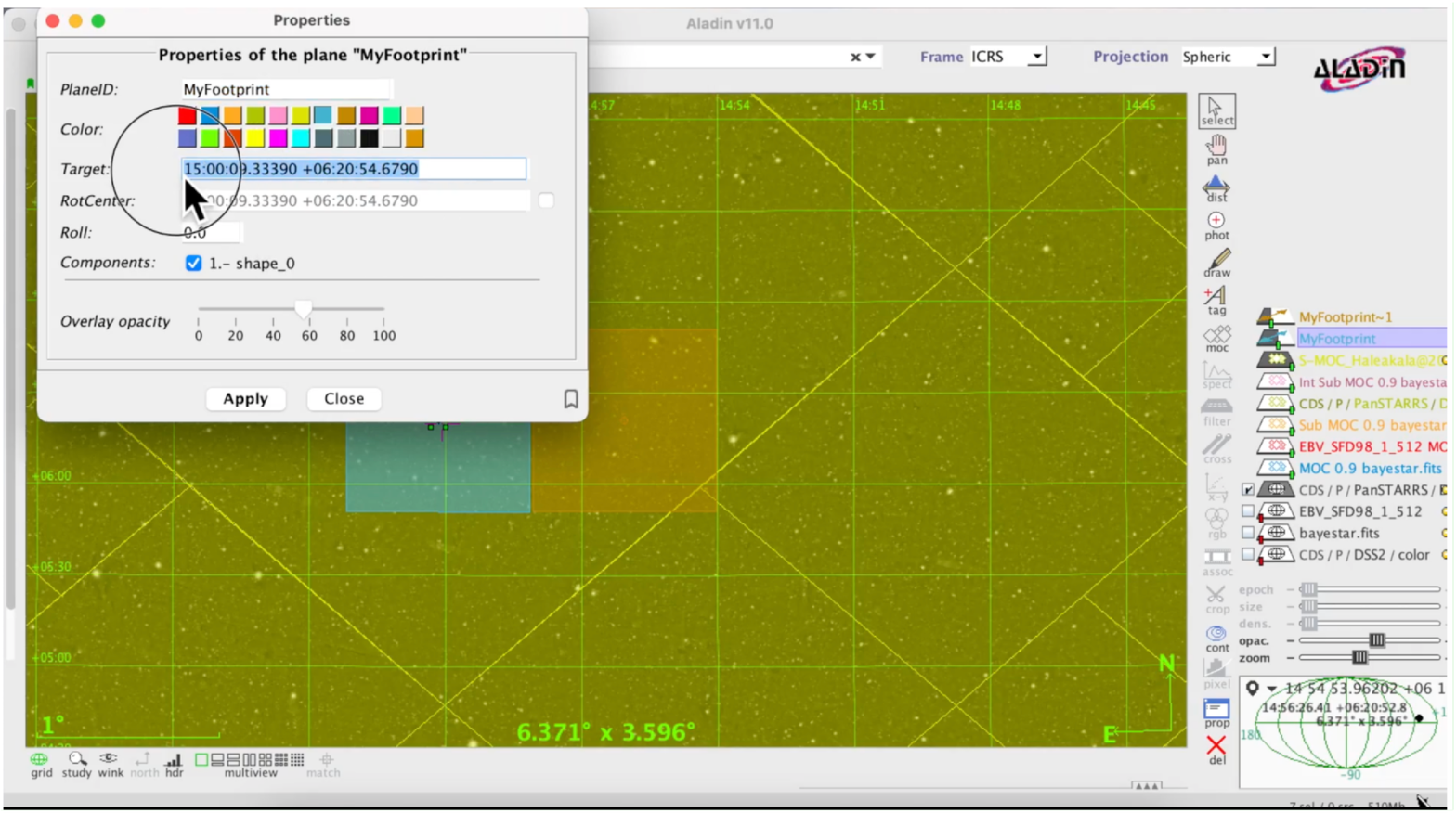}
	\includegraphics[width=0.8\textwidth,]{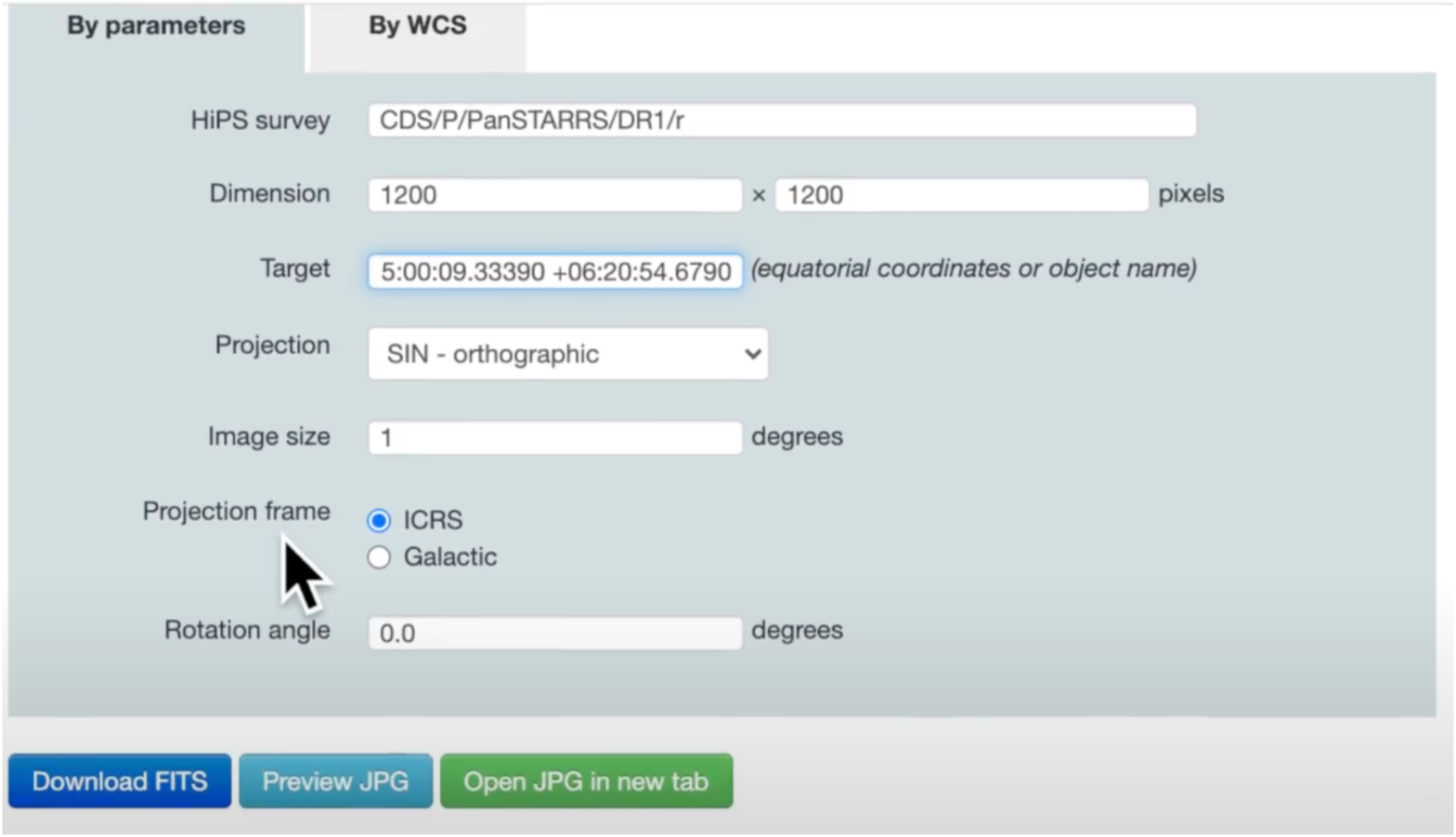}
	\caption{Tiling a sky region and retrieving the images. Two field-of-view footprints (1$^{\circ}$ $\times$ 1$^{\circ}$) are plotted in blue and orange.
	Top: the server selector window in which you can choose/add a field-of-view instrument. 
	Middle: properties window associated with the current field-of-view instrument loaded in the Aladin stack. The box Target shows the sky position centred at the   field-of-view instrument.
	Bottom: web user interface of the hips2fits server. \url{https://alasky.u-strasbg.fr/hips-image-services/hips2fits}}
\end{figure}

The tiling patterns to split the sky areas proportional to the instrument footprints can also be generated using the Aladin Desktop functionality (from \textbf{File $\Rightarrow$ Load Instrument FoV $\Rightarrow$ Server selector}) or dedicated packages; GWsky$\footnote{\url{https://github.com/ggreco77/GWsky}}$,
teglon$\footnote{\url{https://github.com/davecoulter/teglon}}$,
gwemopt$\footnote{\url{https://github.com/mcoughlin/gwemopt}}$,
GW-Localization-Tiling$\footnote{\url{https://github.com/kauii8school/GW-Localization-Tiling}}$, sky\_tiling$\footnote{\url{https://github.com/shaonghosh/sky_tiling}}$, to mention a few.
Three screenshots taken from the video tutorial associated with the present paper compose Figure \ref{fig:tiling_retrieving}. The server selector window, in which the instrument field-of-view can be chosen/added, is shown in the top panel. Two field-of-view footprints are plotted in blue and orange (1$^\circ$ $\times$ 1$^\circ$). The sky position centred at the field-of-view instrument can be taken opening the properties window associated with the current field-of-view instrument loaded in the Aladin stack. That is depicted in the middle panel. Finally the cut-out images can be obtained by filling out the form of the hips2fits web server as shown in the bottom panel.

Note, thumbnails for a list of targets or coordinates are also readily available through the submission of queries from dedicated Python scripts.

 \subsection{Updating Sky localizations}
 \label{subsec:updating}
 Gravitational-wave sky localizations are distributed for a single event increasing accuracy and consequently the computational time  \cite{gw150914_mma}. The sequence of LIGO/Virgo alerts disseminated
 through the Gamma-ray Coordinates Network (GCN), during the first Open Public Alert, are described in the LVC User Guide in the Section Alert Timeline$\footnote{\url{https://emfollow.docs.ligo.org/userguide/index.html}}$.
 
\begin{figure}
\center
	\includegraphics[width=0.8\textwidth, ]{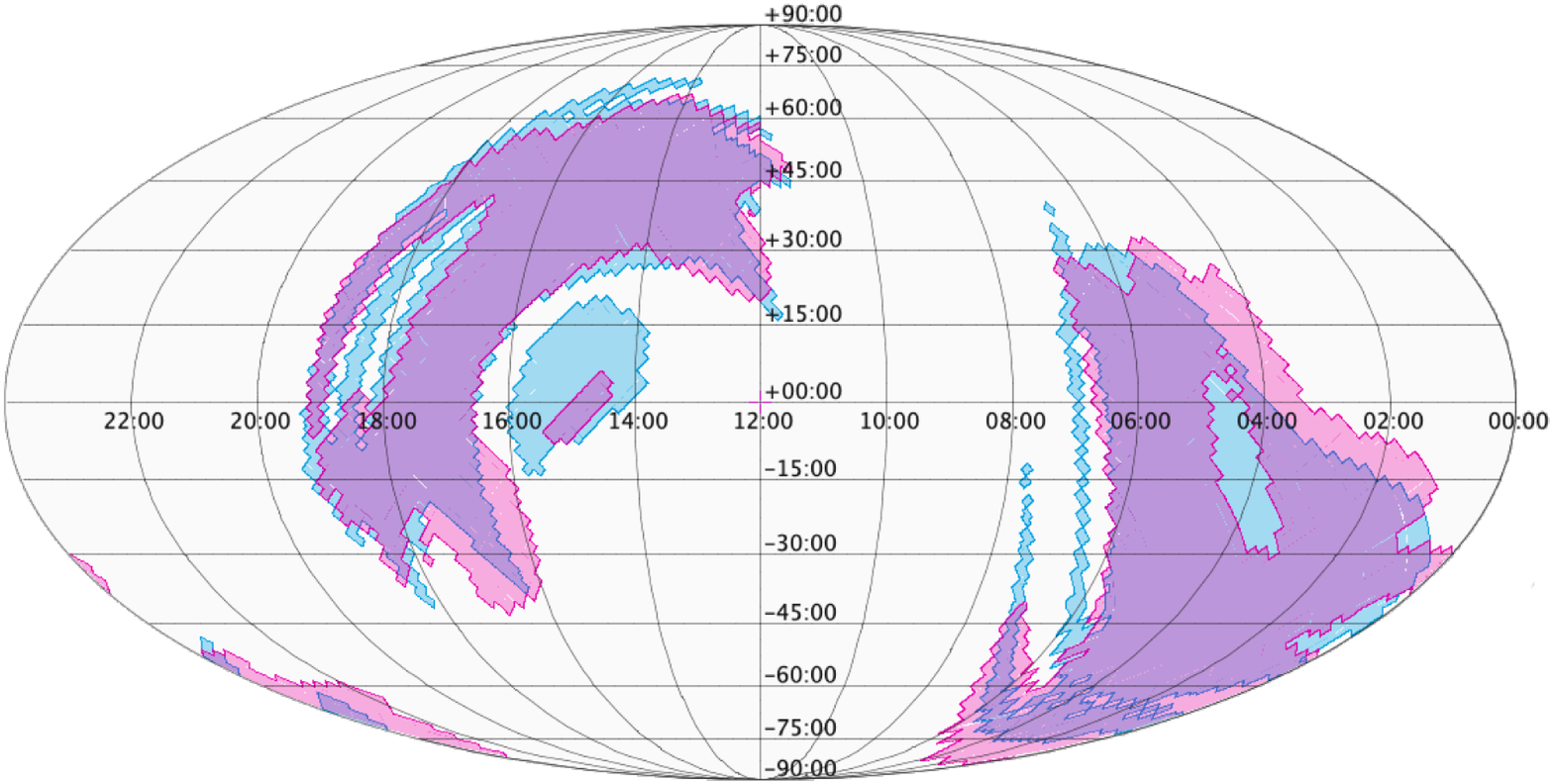}
	\caption{Sky localizations of GW190425. Low-latency sky localization, in light blue, overlapping the sky localization published in GWTC-2, in magenta.}
\label{fig:comparison}
\end{figure}

 Figure \ref{fig:comparison} shows the 90$\%$ confidence region from the low-latency sky localization algorithm BAYESTAR in blue, as previously shown  in the top panel of Figure \ref{fig:skymap_operations_1}, overlapping the sky localization published in GWTC-2 (The second Gravitational Wave Transient Catalogue) \cite{GWTC2}, in magenta.
 The shaded areas, useful for quick visual comparisons, are obtained using as drawing methods \textbf{perimeter} and \textbf{fill} in Properties window: right-click the MOCs in the Aladin stack
and select \textbf{Properties} from the contextual menu.
 
\subsection{Encoding time information in (visibility) MOCs}
Operationally, the addition of time information in a spatial MOC can be achieved using the \texttt{from\_spatial\_coverages} method in \texttt{mocpy} or managing the \textbf{Properties} window in the Aladin Desktop. This is accomplished by selecting the menu bar the item called \textbf{Coverage $\Rightarrow$ Generate a Space-Time MOC based on $\Rightarrow$ The selected Space MOC}.

The resulting new data structure is indicated with STMOC (Space and Time MOC). While a detailed description of STMOC is beyond the scope of this paper, we provide a brief outline based on \cite{fernique21}. The temporal coverage (TMOC) of a data set can be expressed using a discrete time axis where the unit element has a duration of 1 microsecond. Using the same hierarchical approach as the spatial MOCS, the time axis is organised into orders of increasing temporal duration, with each order being a factor of two longer than the previous one. Using 62 orders thus enables temporal coverage spanning $\sim$73000 years (from JD=0) at 1 microsecond resolution. STMOC is the combination of TMOC and the spatial MOC, using a two-dimensional interleaving method for combining the indices and keeping the spatial and temporal coverage resolutions independent.

In the associated tutorial, we make use of the implementation of STMOC in \texttt{mocpy} to simultaneously filter a list of transient candidates according to the spatial and temporal information encoded through STMOC.

\section{Summary and further developments}
Recently, discussions  on telescope coordination have been held by the IAU (International Astronomical Union) Executive Committee Working Group on Coordination of Ground and Space Astrophysics organised, with the support of the Kavli Foundation.  As a result of a Kavli-IAU Workshop\footnote{\url{https://www.iau.org/news/announcements/detail/ann20027/}} a white paper was written summarising the discussions that took place and a set of recommendations and best practises for the community \cite{iau2020}. The development of standards adopted by the community are critical to ensure that heterogeneous transient/multi-messenger systems can communicate efficiently and effectively at a degree of cooperation not currently practised. 

The success of recent visualisation packages such as The Gravitational Wave Treasure Map deserves to be mentioned in this framework \cite{treasure_map}. 
It aims to efficiently organise the multi-messenger effort provided by worldwide collaborations (see instruments reporting$\footnote{\url{http://treasuremap.space/}}$) in order to minimise unnecessary overlap.

In this paper, we have shown various applications for managing the 2D sky maps from source localizations of  gravitational-wave transient events when the enclosed probability areas are encoded with MOC data structures. We have demonstrated that the MOC data structure to plan multi-messenger observations proves to be a very flexible and interoperable methodology to satisfy many research demands in multi-messenger science (data comparison, data access and data visualisation).

In the Python implementation of MOC, excellent performances are obtained thanks to the core functions written in Rust programming language.
Visualisation and operations of MOCs are fully integrated in the VO standard and traditional tools as Aladin Desktop and Aladin Lite.
Dedicated software are also developed to share proprietary images or footprints in the HiPS/MOC data structure$\footnote{\url{https://aladin.u-strasbg.fr/hips/HipsIn10Steps.gml}}$ in order to apply the FAIR (Findable, Accessible, Interoperable,  Reusable) principles in the context of the IVOA ecosystem. 

In more detail, we have investigated an interoperable approach to determine the observability of MOC cells  based on the geographic coordinates of ground-based observatories. Before identifying
the visibility MOC area, the sky localizations can be processed in order to optimise follow-up strategies to increase the chance to identify the electromagnetic counterpart of a gravitational-wave event.
Here we have simulated two basic examples: removing or identifying the highest extinction areas  and collecting information querying VO data providers to search for reference image catalogues. 
In the first operation, we perform a subtraction between the initial gravitational wave sky localization and a dust Galactic map. In the second operation, the subtracted sky map is intersected with the  reference image catalogue chosen from the Aladin data trees. The computation speed (generally a few milliseconds, see Section 1.6.4 in \cite{moc1.1}) is exceptionally high for the MOC operations. This aspect makes them  more efficient than using polygons to confine complex and irregular sky regions.

The visibility of the resulting sky map is quantified from three astronomical sites composing our ideal multi-messenger network. 

Finally, we also encoded  to the MOC map the temporal information adopting the STMOC data structure. An observation campaign that makes use of  visibility (ST)MOCs can easily quantify the observation time between the various observers as soon as an alert has been launched and reshape it in real time 
meeting the current observational requirement.

The complete code is reported in a jupyter notebook  which indicates all of the Python modules necessary for this analysis.

More convenient approaches, especially for instruments with relatively small field of view, can exploit  3D sky maps \cite{3dskymap} ranking galaxies inside the 90\% of  volume  with astrophysical prescriptions to trace potential host galaxies for binary mergers that contain a neutron star \cite{3d_supplementary}, \cite{gehrels}, \cite{antolini}, \cite{Arcavi_2017}, \cite{rana} \cite{ducoin}, \cite{artale}. 
These approaches are highly sensitive to the completeness of  catalogues \cite{coughlin18}. Dedicated interactive web pages have been published to provide access to  galaxy lists such as:
MANGROVE (Mass AssociatioN for GRavitational waves ObserVations Efficiency)$\footnote{\url{https://mangrove.lal.in2p3.fr/index.php}}$,
NED Gravitational Wave Follow-up (GWF) Service$\footnote{\url{https://ned.ipac.caltech.edu/uri/NED::GWFoverview/}}$ and 
HOGWARTs (Hunt Of Gravitational Wave Areas for Rapid Transients)$\footnote{\url{https://github.com/Lanasalmon/HOGWARTs}}$.
 
One of the possible future developments of this work may be to enable the creation of MOC maps based on 3-D gravitational-wave sky localizations. The work presented here has been based on the 2-D probability maps, but this can potentially be extended to 3-D via use of additional information in the LIGO-Virgo-KAGRA sky maps. In future work we will explore the use of additional layers dedicated to measure the conditional distance distribution along a line of sight and extract the galaxies within the 90\% credible volume. The idea is to convert the galaxy list "from 3d sky map" in a catalogue HiPS.

\begin{table}[ht]
\small
\caption{MOC visibility areas [deg${^2}$]. } 
\centering 
\begin{tabular}{c c c c } 
\hline\hline 
Obs Time (UTC) & Haleakala & Paranal & SSO   \\ [0.5ex] 
\hline 
\texttt{2019-04-25 08:18:05.0} & 2567 & 2038 & -- \\ 
\texttt{2019-04-25 10:18:05.0} & 3989 & --   & 126 \\
\texttt{2019-04-25 12:18:05.0} & 4334 & --   & 767 \\
\texttt{2019-04-25 14:18:05.0} & 3711 & --   & 1500 \\
\hline 
\end{tabular}
\label{table:areas} 
\end{table}

\begin{table}[ht]
\scriptsize
\caption{MOC visibility intersection areas [deg${^2}$]. } 
\centering 
\begin{tabular}{c c c c c } 
\hline\hline 
Obs Time (UTC) & Haleakala $\cap$ Paranal  $\cap$ SSO & Haleakala $\cap$ Paranal & Haleakala $\cap$ SSO &  Paranal $\cap$ SSO  \\ [0.3ex] 
\hline 
\texttt{2019-04-25 08:18:05.0} & --  & 657  & --   & --\\ 
\texttt{2019-04-25 10:18:05.0} & --  & --   & 31   & --\\
\texttt{2019-04-25 12:18:05.0} & --  & --   & 761  & --\\
\texttt{2019-04-25 14:18:05.0} & --  & --   & 1450 & --\\
\hline 
\end{tabular}
\label{table:intersection} 
\end{table}


\paragraph{Acknowledgements}
The research leading to these results has received funding from the European Union’s Horizon 2020 Programme under the AHEAD2020 project (grant agreement n. 871158). This work has been partly supported by the ESCAPE project (the European Science Cluster of Astronomy \& Particle Physics ESFRI Research Infrastructures) that has received funding from the European Union’s Horizon 2020 research and innovation programme under the Grant Agreement n. 824064.


\bibliography{mybibfile}

\end{document}